\newif\iftodos
\todostrue

\newif\ifpreprint
\preprintfalse

\newif\ifanon
\anonfalse

\newif\ifrevisions
\revisionsfalse

\newif\ifbiblatex
\biblatextrue

\newif\ifpublic
\publictrue

 \preprinttrue

 \ifpublic
    \revisionsfalse
    \todosfalse
\fi

\ifpreprint
\documentclass[sigconf,natbib=false,screen,authorversion]{acmart}
\else
\biblatexfalse
\documentclass[sigconf]{acmart}
\fi

\usepackage{tabularx}
\usepackage{subcaption}

\newcommand{\rev}[1]{{#1}}
\newcommand{\todo}[1][]{{}}
\newcommand{\drtodo}[1][]{{}}
\newcommand{\rstodo}[1][]{{}}

\newcommand{\quoteformat}[1]{\emph{#1}}
\newcommand{\inlinequote}[2]{``\quoteformat{#1}'' (P#2)}
\newcommand{\blockquote}[2]{%
    \begin{quote}\quoteformat{#1} (P#2)\end{quote}
}

\makeatletter
\newcommand{\inlinesection}{\@ifstar{\@inlinesectionstar}{\@inlinesectionnostar}}
\newcommand{\@inlinesectionstar}[1]{\textbf{\textit{#1}}}
\newcommand{\@inlinesectionnostar}[1]{\@inlinesectionstar{#1}}
\makeatother

\usepackage{datetime}
\newdateformat{monthdate}{%
  \monthname[\THEMONTH], \THEDAY}

\newcommand{\subs}[1]{\emph{(#1)}}

\newcommand{\vizurl}{\href{https://tools.auditing-ai.com}{tools.auditing-ai.com}}
\newcommand{\github}{\href{https://github.com/ryansteed/oat-analysis}{github.com/ryansteed/oat-analysis}}

\newcommand{\nInterviews}{27}
\newcommand{\nParticipants}{35}
\newcommand{\nParticipantOrgs}{24}

\newcommand{\nToolsRecent}{102} %
\newcommand{\nToolsTargeted}{181} %
\newcommand{\nToolsSeed}{79} %
\newcommand{\nToolsResample}{9} %
\newcommand{\nToolsInitial}{143} %

\newcommand{\nTools}{435}
\newcommand{\nOrgs}{347}
\newcommand{\nToolsCB}{270}
\newcommand{\nOrgsCB}{202}
\newcommand{\nOrgsCBEmployees}{173}
\newcommand{\nOrgsCBFirms}{132}
\newcommand{\nOrgsCBFirmsRev}{91}
\newcommand{\nOrgsCBFirmsPrivate}{112}
\newcommand{\nOrgsCBFirmsPrivateFunding}{48}

\newcommand{\nHarms}{48}
\newcommand{\nStandards}{206}
    \newcommand{\nStandardsGoal}{86}
    \newcommand{\nStandardsAssessment}{49}
    \newcommand{\nStandardsParticipatory}{5}
\newcommand{\nTransp}{12}
\newcommand{\nData}{47}
\newcommand{\nPerf}{129}
    \newcommand{\nPerfAcc}{71}
    \newcommand{\nPerfExp}{36}
    \newcommand{\nPerfQual}{3}
    \newcommand{\nPerfFair}{30}
\newcommand{\nComm}{8}
    
\newcommand{\nAdvocacy}{14}

\newcommand{\osPropTotal}{77.9\%} %

\newcommand{\osPropData}{68.1\%}
\newcommand{\osPropTransp}{25.0\%}

\newcommand{\notprofitPropHarms}{79.2\%} %

\newcommand{\profitNumTransp}{9} %

\newcommand{\profitPropPerf}{48.1\%}

\AtBeginDocument{%
  }

\copyrightyear{2025}
\acmYear{2025}
\setcopyright{cc}
\setcctype{by}

\acmConference[CHI '25]{CHI Conference on Human Factors in Computing Systems}{April 26-May 1, 2025}{Yokohama, Japan}
\acmBooktitle{CHI Conference on Human Factors in Computing Systems (CHI '25), April 26-May 1, 2025, Yokohama, Japan}\acmDOI{10.1145/3706598.3713301}
\acmISBN{979-8-4007-1394-1/25/04}

\ifpreprint
\setcopyright{none}
\renewcommand\footnotetextcopyrightpermission[1]{} %
\acmConference[Preprint]{}{\monthdate\today}{\THEYEAR}
\fi

\ifbiblatex
\RequirePackage[
  datamodel=acmdatamodel,
  style=acmnumeric,
  natbib=true,
  sortcites,
  ]{biblatex}
\fi

\begin{document}

\title{Towards AI Accountability Infrastructure:  Gaps and Opportunities in AI Audit Tooling}

\author{Victor Ojewale}
\affiliation{%
  \institution{Brown University}
  \city{Providence}
  \state{Rhode Island}
  \country{USA}
}

\author{Ryan Steed}
\affiliation{%
  \institution{Carnegie Mellon University}
  \city{Pittsburgh}
  \state{Pennsylvania}
  \country{USA}}

\author{Briana Vecchione}
\affiliation{%
 \institution{Data \& Society}
 \city{New York City}
 \state{New York}
 \country{USA}}

\author{Abeba Birhane}
\affiliation{%
  \institution{Mozilla Foundation \textit{and}}
  \institution{Trinity College Dublin}
  \city{Dublin}
  \country{Ireland}}

\author{Inioluwa Deborah Raji}
\affiliation{%
  \institution{Mozilla Foundation \textit{and}}
  \institution{University of California, Berkeley}
  \city{Berkeley}
  \state{California}
  \country{USA}
}

\renewcommand{\shortauthors}{Ojewale et al.}

\begin{abstract}

Audits are critical mechanisms for identifying the risks and limitations of deployed artificial intelligence (AI) systems. However, the effective execution of AI audits remains incredibly difficult, and practitioners often need to make use of various tools to support their efforts. Drawing on interviews with {\nParticipants} AI audit practitioners and a landscape analysis of {\nTools} tools, we compare the current ecosystem of AI audit tooling to practitioner needs. While many tools are designed to help set standards and evaluate AI systems, they often fall short in supporting accountability. We outline challenges practitioners faced in their efforts to use AI audit tools and highlight areas for future tool development beyond evaluation---from harms discovery to advocacy. We conclude that the available resources do not currently support the full scope of AI audit practitioners' needs and recommend that the field move beyond tools for just evaluation and towards more comprehensive infrastructure for AI accountability.

\end{abstract}



\maketitle

\section{Introduction}

Despite increasing policy enthusiasm,\footnote{AI audits have been featured in several recent U.S. congressional  bills~\citep{clarke_algorithmic_2019, lenhart_federal_2023} and state efforts~\citep{perrigo_california_2023,mendelson_stop_2021}, and the practice is regularly mentioned in AI governance proposals internationally~\cite{galindo_overview_2021}, from the E.U. Digital Services Act to a municipal hiring bill passed in New York City \citep{cumbo_local_2021}.
} the execution of effective \emph{AI audits} remains practically difficult. Often defined as independent evaluations of the performance, fairness, legality, or safety of deployed AI systems, the maturity of the audit ecosystem in the technology sector %
lags far behind %
other industries such as finance and healthcare~\citep{raji_closing_2020, raji_outsider_2022}.\footnote{This dearth of generalized AI audit guidance is remedied only partially by recent efforts from government advisory bodies like the U.K.'s Information Commissioner's Office (ICO)~\cite{information_commissioners_office_annex_2023}, the U.S. National Institute of Standards and Technology~\cite{tabassi_ai_2023} and others \citep{office_of_science_and_technology_policy_blueprint_2022}.}
AI audits are often inconsistent and unreliable \citep{ryan-mosley_why_2023}, and the lack of access and visibility to many AI %
systems leaves auditors without the information needed to make adequate and truly independent assessments \citep{holstein_improving_2019, terzis_law_2024}.

In the face of these challenges, practitioners often rely on tools---software, frameworks, and other resources---to support their AI audit work. Past research in human-computer interaction (HCI) and social computing has developed and studied a host of fairness, explainability, and other toolkits that inform such evaluations \citep{bellamy_ai_2018, smith-renner_no_2020, kaur_interpreting_2020, bertrand_selective_2023, deng_understanding_2023, woodruff_qualitative_2018, brown_toward_2019, madaio_co-designing_2020, devos_toward_2022, lee_landscape_2021, holstein_improving_2019, deng_exploring_2022, amershi_modeltracker_2015, wong_seeing_2023, ehsan_human-centered_2020}. Governments across the globe %
are developing their own tools and making use of existing resources as part of their enforcement regimes for AI governance \citep{kaye_risky_2023}.\footnote{Examples include the U.S. AI Safety Institute's Inspect, the U.S. National Institute of Standards and Technology's ARIA \& Dioptra, Singapore's AI Verify, and the U.S. National Science Foundation's Artificial Intelligence Research Resource (NAIRR) Pilot.}
In the U.S. alone, several recently proposed ``AI innovation'' bills are explicitly geared towards investing in tooling and resource development for AI auditing.\footnote{%
Examples include the ``CREATE AI Act of 2023'', ``VET Artificial Intelligence Act'', ``Promoting United States Leadership in Standards Act of 2024'', ``TEST AI Act of 2023'', ``Artificial Intelligence Research, Innovation, and Accountability Act of 2023'', ``Artificial Intelligence Public Awareness and Education Campaign Act'', and ``Future of Artificial Intelligence Innovation Act of 2024''.}
Similarly, in the E.U., enforcement reports for the Digital Services Act %
emphasize the importance of AI audit tooling for effective oversight enforcement~\citep{klinger_delegated_2023}.

Despite these developments, research has yet to properly map, taxonomize, and understand the full scope of tooling needed to meaningfully support AI audit practitioners.
HCI research has identified many practical challenges facing practitioners \citep{holstein_improving_2019, costanza-chock_who_2022, brown_toward_2019, madaio_co-designing_2020, devos_toward_2022} and valuable recent work critically examines some of the toolkits involved
\citep{lee_landscape_2021, deng_exploring_2022, wong_seeing_2023, berman_scoping_2024}.
However, the auditing process involves more than just a performance analysis of the AI product or model. A thorough evaluation alone is not sufficient to hold key stakeholders responsible for system-wide behavior \citep{goodman_ai_2022, raji_outsider_2022}---auditors also need tools that support key components of the accountability process, including target identification, standardization of practice, communication, and advocacy \citep{costanza-chock_who_2022}.

In this study, we compare audit practitioners' tooling needs to the current landscape of AI audit tooling to understand challenges to accountability and potential opportunities for expanding the scope of HCI research and tool development.
\begin{enumerate}
    \item \label{item:rq1} \textbf{RQ\ref{item:rq1}: What tools do practitioners need to conduct AI audits?}
    \item \label{item:rq2} \textbf{RQ\ref{item:rq2}: What tools currently exist to support AI audit work?}
    \item \label{item:rq3} \textbf{RQ\ref{item:rq3}: Do existing tools support practitioners' needs?}
\end{enumerate}
To investigate RQ\ref{item:rq1}, we interviewed {\nParticipants} AI audit practitioners---employed at {\nParticipantOrgs} organizations, including tech companies, startups, government agencies, non-profits, consulting firms, and academic institutions---about the tools they use and how they support the audit process. To investigate RQ\ref{item:rq2}, we conducted a landscape analysis of {\nTools} existing audit tools (Fig.~\ref{fig:taxonomy}).\footnote{Note that our sample is not an exhaustive list of all AI audit tools.} In the interviews, we asked practitioners about gaps they encountered and compared their responses to the current tooling landscape (RQ\ref{item:rq3}).

\newcommand{\taxonomyDisclaimer}{Tools may be used in multiple stages.}
\begin{figure*}[!t]
  \centering
  \includegraphics[width=\linewidth]{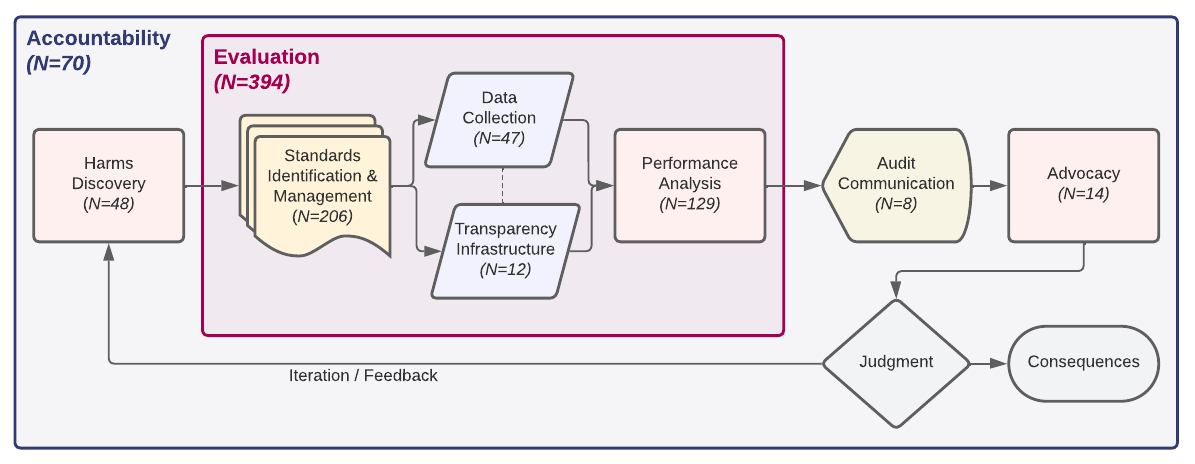}
  \caption{Stages of the tool-supported audit process surfaced in our survey of AI audit tooling. We taxonomize tools by the stage of the AI audit process in which they are used. \taxonomyDisclaimer{}} %
  \Description{Flowchart with seven elements connected by arrows in the following order: Harms Discovery, Standards Identification \& Management, Data Collection / Transparency Infrastructure (grouped together), Performance Analysis, Audit Communication, Advocacy. Advocacy then points to a decision element called `Judgment`, which leads to `Consequences'. `Judgment' also leads back to Harms Discovery through an arrow labeled `Iteration/Feedback'. All the elements are enclosed in a box labeled `Accountability'. The elements from Standards Identification to Performance Analysis are in a box labeled `Evaluation'.}
  \label{fig:taxonomy}
\end{figure*}

We find that while there currently exist many tools to support audit work, particularly for evaluating AI systems and managing standards, these tools often fell short of helping auditors achieve accountability in practice. Practitioners found some tools---such as open source tools for data collection---empowering, %
but tools for tasks beyond evaluation, such as discovering harms, communicating audit results, and advocating for subsequent changes, were much less common. 
The practitioners we interviewed often adapted existing tools or built their own from scratch to relieve the tedious or difficult tasks in their particular audit workflows. Auditors envisioned tools to help them access high-quality, uncompromised data, apply consistent and holistic standards and methods, and ensure audit integrity. %

Our results identify challenges for every stage of the AI audit process %
as well as %
opportunities for HCI researchers, policymakers, and practitioners to pursue an alternative vision for tool development that supports rigor, inclusion, %
independence, and accountability. We conclude with a summary of research and design opportunities for the HCI community and other stakeholders that could help push the landscape of AI audit tools beyond evaluation and towards infrastructure for meaningful accountability.

\section{Related work}

\inlinesection{AI auditing.}
The practice of AI auditing, or algorithm auditing---a term first formally proposed in 2014 by \citet{sandvig_auditing_2014} to describe methods for detecting discrimination in online platforms---has expanded over the last decade \citep{vecchione_algorithmic_2021}. Researchers have since used the term to encompass not just field studies for detecting discrimination with causal estimation~\citep{metaxa_auditing_2021} but also any kind of independent assessment of an automated or data-defined system \citep{digital_regulation_cooperation_forum_auditing_2022, raji_closing_2020}. Early audit studies of facial recognition systems and criminal risk assessments, \citep{angwin_machine_2016, buolamwini_gender_2018}, for example, resulted in widespread advocacy and, at times, even changes to or recalls of the audited systems \citep{raji_actionable_2022, sherwin_facebook_2019, spinks_contemporary_2020, sheard_banning_2021}. However, not all audits succeed in holding system builders and operators accountable \citep{goodman_ai_2022, watkins_governing_2021, birhane_ai_2024}.
\newcommand{\defAudit}{``any independent assessment of an identified audit target via an evaluation of articulated expectations with the implicit or explicit objective of accountability''}

Following \citet{birhane_ai_2024}, we use the term \emph{AI audit} (Def.~\ref{def:audit}) to refer to \defAudit{}. Accountability is used here in the legal-political sense to mean consequential judgment of a systems' behaviors and downstream impacts~\cite{bovens_analysing_2007}. Consequential judgment distinguishes an audit from a simple evaluation or assessment (Fig.~\ref{fig:taxonomy}).

More recently, corporations and policymakers have focused on what we refer to as \emph{internal} AI audits (Def.~\ref{def:auditor}) conducted by teams of employees or contractors separate from the product and engineering teams
\citep{european_parliament_regulation_2016, raji_outsider_2022}. These internal audits, afforded access in cooperation with audit targets, may enable accountability if designed properly \citep{raji_closing_2020, desai_trust_2017}, but they may also result in false assurances (``audit washing'')~\citep{raji_outsider_2022, goodman_ai_2022} or foreclose on key remedies such as abandonment or disgorgement (Def.~\ref{def:abandonment-disgorgement}) \citep{sloane_silicon_2022, li_algorithmic_2022, johnson_fall_2024}. As a result, there is an %
important role for \emph{external} AI audits (Def.~\ref{def:auditor}): investigations conducted by civil society, journalists, lawyers, regulators, and other third-party actors. These external audits are typically voluntary research studies and investigations into %
deployed AI systems. However, as our findings reveal, external auditors faced significant hurdles to accountability, including lack of access %
to the systems they aimed to evaluate. %
The experiences and challenges of those doing external audit work has not been the primary focus of HCI research. Our work provides %
references for understanding their tooling and resource needs~\cite{bandy_problematic_2021, birhane_ai_2024, costanza-chock_who_2022}.

While these prior studies define the practice of AI auditing and some of the components of accountability, no study yet examines the full range of \emph{tools} and technical infrastructure used to support AI auditing and accountability. 

\inlinesection{Infrastructure \& accountability.}
A rich field of literature in science, technology and society (STS) describes infrastructure as not merely physical or technological, but also as holding social and relational power \cite{winner_artifacts_1980, star_ethnography_1999}.
Infrastructure is ``something that other things `run on''' \citep{lampland_standards_2009}: the invisible and axiomatic basis of tools, standards and frameworks that uphold and shape more complex existing processes, services, and engineered artifacts.
Embedded in social systems, technological infrastructures encode standards, norms, and guidelines for social organization %
\citep{verbeek_materializing_2006, lampland_standards_2009}.

Accountability also requires infrastructure. Across various industries, the common accountability practice of auditing has long relied on rituals, organizational processes and tools in order to make consistent, inter-operable and reliable judgments~\cite{power_audit_1999}, as well as provide broader access to a larger range of stakeholder participants in the audit process~\cite{birhane_power_2022}. Audit tooling, then, represents not only a mechanism for maintenance and consistency of audit integrity but also a key capacity-building intervention to lower the barriers for broader participation.

\inlinesection{Past work on AI audit tools.} 
Recent research in human-computer interaction (HCI), social computing, and cooperative design documents the experiences of practitioners evaluating AI and the challenges they face~\citep{costanza-chock_who_2022, holstein_improving_2019, brown_toward_2019, madaio_co-designing_2020, devos_toward_2022, lee_landscape_2021}. For example, \citet{holstein_improving_2019} documents the \emph{practical} and \emph{technical} difficulties faced by internal auditors of ML systems trying to identify and improve fairness, while other studies have examined the \emph{organizational} challenges and barriers these practitioners face \citep{rakova_where_2021, madaio_co-designing_2020, widder_its_2023, selbst_fairness_2019, costanza-chock_who_2022}. 
Several HCI studies have since specifically examined the ways AI audit practitioners make use of various tools to address these issues \citep{deng_exploring_2022, lee_landscape_2021, amershi_modeltracker_2015}.

While some studies of AI audit tools focus on performance analysis \citep{amershi_modeltracker_2015, harvey_gaps_2024} or user-driven grassroots auditing \citep{devos_toward_2022, deng_understanding_2023}, most HCI studies focus specifically on tools for assessing fairness and interpretability \citep{bellamy_ai_2018, smith-renner_no_2020, kaur_interpreting_2020, bertrand_selective_2023, woodruff_qualitative_2018, madaio_co-designing_2020, lee_landscape_2021, deng_exploring_2022}. \citet{lee_landscape_2021}, for example, compare six prominent open source fairness toolkits along various criteria related to practitioners' needs. \citet{deng_exploring_2022} documented the ways practitioners learn about and use two prominent fairness toolkits, AI Fairness 360 and Fairlearn \citep{weerts_fairlearn_2024}. And in a survey of 152 audit practitioners, \citet{costanza-chock_who_2022} found that 62\% of practitioners used existing tools like AI Fairness 360, Scikit Fairness, or Parity, though only 7\% of respondents used a standardized framework for their overall audit protocol.

However, less work examines tools for assessing harms beyond fairness, including refusal of medical %
services \citep{waldman_how_2024}, privacy violations, and other types of harm \citep{shelby_sociotechnical_2023}.
And fairness toolkits do not typically support other necessary steps of an audit such as data collection. 
Our study aims to expand HCI research by looking beyond fairness evaluation toolkits to examine other kinds of tools involved in AI audits, such as tools for basic performance analysis or data collection.

We also build on more general critiques of ``responsible AI'' toolkits. \citet{wong_seeing_2023}, for example, examine documentation from 27 toolkits for ``AI Ethics'', finding that these resources employ a narrow technical framing that fails to involve more diverse stakeholders or reckon with the non-technical dimensions of AI ethics work; \citet{kaye_risky_2023} survey and critiques the tools currently used for AI governance by governments across the globe; and \citet{berman_scoping_2024} call for more evaluations of the effectiveness of responsible AI tools.

\newcommand{\defAuditTool}{software, interfaces, code, benchmarks, frameworks, and other artifacts used by auditors in the AI audit process}

As yet, however, %
a comprehensive survey of auditors' practical needs relative to the landscape of available tools is lacking.
We extend these analyses and %
lay out a more complete view on what we refer to as \emph{AI audit tools}: 
{\defAuditTool} (Def.~\ref{def:audit-tool}).\footnote{%
    We include in this definition tools that may also be used for other ``responsible AI'' efforts, such as internal benchmarking, that do not meet our criteria for an audit (Def~\ref{def:audit}). Auditing is an institutional arrangement---selecting the right tools does not guarantee operational independence, for example.
}

\section{Methodology}
To better understand the kinds of tools auditors use %
and where those tools fall short, %
we conducted {\nInterviews} semi-structured interviews with {\nParticipants} auditors across {\nParticipantOrgs} organizations employing internal and external AI auditors.
We use the term %
AI broadly %
to include tools applied to any AI-advertised product or model, including automated decision systems (ADS), algorithmic recommendation systems, large machine learning base models, generative AI products and more (see Fig.~\ref{fig:target-count-second}).
In parallel, to better understand existing tools, %
we %
curated a dataset of {\nTools} tools designed or used for AI auditing and developed a taxonomy of the audit tool landscape based on our findings.

\subsection{Interview methodology}
We conducted {\nInterviews} interviews with a total of {\nParticipants} audit tool builders and practitioners, representing diverse backgrounds such as engineering, law, journalism, advocacy, policy, and academic research across %
North America ($N=22$) and Europe ($N=5$) (Table~\ref{tab:sample}). %
These practitioners have all participated in internal or external audit work; many have also built tools for AI auditing.
We used purposive sampling and snowball sampling methods to recruit participants. We began by contacting practitioners in our professional networks who had conducted notable AI audit work and were active in AI audit communities. Occasionally, participants were referred to us by a colleague or professional contact at another organization. Our sample encompassed both \emph{internal} and \emph{external} auditors employed by for-profit tech companies, AI evaluation startups, research %
and civil society non-profits, universities, and government agencies.

\begin{table*}[!t]
    \centering
    \caption{Participants' organizations and titles at the time of interview. Some titles are summarized for anonymity. Participants in the same interview are grouped in parentheses.}
    \begin{tabularx}{\linewidth}{p{11em} X p{10em}}
        \toprule
        Employer & Roles of Interviewees & Participants \\
        \midrule
        Large tech for-profit & Director of Policy Research, VP of Research, Data Science Mgr., Research Eng., Researcher & P5, P8, P11, P14, P19, P27 \\
        Tech startup & Co-Founder, CEO, Chief Scientist & P4, P12 \\
        Government agency & Tech Policy Principal/Mgr./Assoc./Advisor, Research Fellow & (P21, P28-35), P25 \\ 
        University & Assoc./Asst. Professor, Postdoc. Fellow, Data Scientist & P3, P10, P13, P17, P20 \\ 
        Research non-profit & Co-Founder, Director, Research Scientist & P9, P16, P18 \\ 
        Civil society non-profit & Director, Head of Analytics, Statistician, Researcher, Policy Fellow  & P1, P2, P6, P22, P24 \\ 
        Non-profit news org. & Opinion Writer, Data Journalist & P7, P15 \\ 
        Law/consulting for-profit & Policy Director, Mgr., Consultant & (P23, P28), P27 \\ 
        \bottomrule
    \end{tabularx}
    \label{tab:sample}
\end{table*}

Interviews followed a semi-structured format and lasted 30--60 minutes. Our questions centered on 1) the specific tools and methods practitioners built or employed and 2) common obstacles and unmet needs. %
(The full interview protocol is included in Appendix~\ref{app:protocol}.) Participants had the option to remain anonymous and skip questions at their discretion, though none skipped a question. Our protocol was approved by three university IRBs. 
To analyze the interview data, we transcribed each interview and annotated the transcripts with manual codes. Our coding approach followed an inductive methodology, allowing patterns and themes to emerge from the data \citep{wolcott_transforming_1994, glaser_discovery_2017}. We employed a combination of descriptive coding, which captured the content of the interviews, and values coding, which captured the attitudes and beliefs expressed by participants. Through collaborative sessions and memo writing, we organized these codes and related quotes into key insights, presented in \S\ref{sec:results}.

\subsection{Tool Taxonomy}
\label{sec:methods-tools}

\inlinesection{Initial search.}
To taxonomize the landscape of tools available to support AI audit work, we first developed an initial list of tools and tool-building organizations by searching for tools mentioned in a dataset of published audits from academic audit studies, news articles, government reports and frameworks, white papers from civil society organizations, law firm reports, and case files \citep{birhane_ai_2024} as well as existing lists of tools such as \citep{hickock_ethical_2023} (see Appendix~\ref{app:initial-sources} for details). Our initial search was conducted in August 2022. We also included specific tools and sources that were mentioned in our interviews with practitioners.
After we had collected an initial list of {\nToolsInitial} tools, we developed an initial taxonomy by clustering tools into 21 initial categories based on their intended or actual uses in AI audit work. This approach served as a starting point for category development rather than as an exhaustive inventory.

\inlinesection{Theoretical sampling.} 
Next, we expanded our initial set of tools with two kinds of additional theoretical sampling. With targeted keyword searches on English Google and GitHub, we searched explicitly for additional tools in areas where we had fewer examples until theoretically fresh examples of tools ceased to arise.\footnote{In this stage, we aimed for theoretical saturation, in the style of grounded theory \citep{charmaz_constructing_2014}.} Our search queries were descriptors from our initial categories or descriptors used by tools already collected---alone and combined with terms like ``audit tool'' or ``responsible AI'' (see Appendix~\ref{app:methods-tool-searching}). We expanded our list of sources based on our initial taxonomy. (For example, to find tools in our initial ``Participatory'' category, we searched in the proceedings of participatory AI workshops). We also followed links and references in our initial sample of tools to identify additional, similar tools (snowball sampling). With these methods, we added {\nToolsTargeted} tools between August and October 2022, and we continued to update the dataset with {\nToolsRecent} more tools through September 2024. In September 2024, we ran our search queries again for the top-level categories, adding {\nToolsResample} more tools.

These searches surfaced new examples which we used to expand and re-define our categories. %
We iteratively revised our taxonomy twice more to accommodate new examples, integrate findings from the interview study, and clarify or expand our initial categories. We did not explicitly ask interview participants about these categories, but we did incorporate the tools they used and their descriptions of audit tool use while developing the taxonomy. Our final taxonomy groups tools into 30 main categories with 27 subcategories (Table~\ref{tab:taxonomy}) grounded in the properties of the tools we found and shaped by our interviews with and experiences as AI audit practitioners. We sorted these categories into 7 ``stages'' of the tool-assisted audit process (Fig.~\ref{fig:taxonomy}).

This stage of our search was designed to iteratively refine conceptual categories with the goal of generalizing from empirical descriptions to theoretical insights \citep{lee_generalizing_2003}.
So while our search for tools was systematic, it was not exhaustive. %
The sample represented in our dataset does not include every tool that could support AI audits. Likewise, though our sample is designed to represent commonly used and commonly built tools
in a theoretically representative set of categories, 
numeric descriptions of our sample may not reflect the statistical distribution of all AI audit tools used in practice. In particular, our theoretical sampling strategy deliberately over-represents some less common kinds of tooling (such as tools for advocacy). And because our search relied on public materials, we likely over-represent open tooling over proprietary tooling (though our dataset includes both).
As a result, while our dataset reflects a structured and iterative search, it should not be viewed as a comprehensive or perfectly representative sample.

\inlinesection{Landscape analysis.}
To analyze the qualities of tools across our taxonomy, we also manually labeled each tool with several tags describing the tool's documentation and function, including license (open source or not), organization type (for-profit, non-profit, government, or academic), and other characteristics. One author created the labels and at least one other author reviewed each label for agreement. We also supplemented our dataset with funding \& employment data from \href{http://www.crunchbase.com}{Crunchbase} (accessed in September 2023), activity data from \href{https://github.com}{Github} (September 2024), and citation data from \href{https://scholar.google.com}{Google Scholar} API (September 2023).
Detailed methods and additional analysis can be found in Appendix~\ref{app:landscape}, and a full interactive version of our dataset can be viewed at {\vizurl}.\footnote{
    All the code for our analysis and resulting plots---as well as instructions for obtaining supplemental data---can be accessed at {\github}.
}

\section{Results}
\label{sec:results}
While there %
exist many tools for aiding %
AI auditing, practitioners found existing tools inadequate in multiple ways. Practitioners struggled to independently access high-quality data about system behavior, apply consistent and holistic standards and methods, ensure audit integrity, involve affected stakeholders, and collaborate across disciplines.

In particular, though we found many tools for evaluating the performance of AI systems, current tooling did not always help practitioners reach their accountability goals. First, while our survey of AI audit tools ($N=\nTools$), revealed a wide-ranging landscape of resources built by a variety of academic, for-profit, non-profit, and government organizations (Table~\ref{tab:taxonomy}), we primarily surfaced tools for \emph{evaluation}, particularly tools for Standards Identification \& Management ($N=\nStandards$) and Performance Analysis ($N=\nPerf$). Tools for other stages of the audit process crucial to accountability---Harms Discovery ($N=\nHarms$), Audit Communication ($N=\nComm$), Advocacy ($N=\nAdvocacy$), and model/data transparency ($N=\nTransp$)---were much less common in our sample.

\begin{table*}[!t]
\centering
\caption{High-level description of the tool taxonomy categories. (Visit \rev{{\vizurl}} for an interactive visualization).}
\label{tab:taxonomy}
\begin{tabularx}{\linewidth}{ >{\raggedright}p{5em} >{\raggedright}p{14em} c X X }
\toprule
\textbf{Stage} & \textbf{Categories \subs{Subcategories}} & \textbf{$N$} & \textbf{Purpose} & \textbf{Examples} \\ 
\midrule
Harms Discovery & Education / Awareness \subs{community education, visioning}, Incident Reporting \subs{incident databases, intake forms, bug bounties, hotlines}, Target Identification \subs{algorithm visibility} & {\nHarms} & Help auditors identify and prioritize audit targets and harms to investigate. & ACLU Wa.'s \href{https://www.aclu-wa.org/AEKit}{Algorithm Equity Toolkit}, \href{https://incidentdatabase.ai/}{AI Incident Database}, \href{http://algorithmtips.org}{Algorithm Tips} \\ 
\hline

Standards Identification \& Mgmt. & Goal Articulation \subs{principle statements, standards formulation}, Self-Assessment \subs{checklists, grading}, Documentation \subs{single stage, continuous, licenses}, Regulatory Awareness \subs{discovery, monitoring}, Methods Design, Participatory Standards-Setting & {\nStandards} & Help auditors identify and formulate principles and norms to guide their investigations. & \href{https://ethical.institute/rfx.html}{AI-RFX Procurement Framework}, Microsoft's \href{https://www.microsoft.com/en-us/research/project/ai-fairness-checklist/}{AI Fairness Checklist}, Model Cards \citep{mitchell_model_2019}, Queensland's Community Engagement Toolkit \citep{queensland_government_community_2017}, \href{https://learn.microsoft.com/en-us/azure/architecture/guide/responsible-innovation/community-jury/}{Community Jury} \\ 
\hline

Transparency Infrastructure & Structured/API Access, Secure \& Private Sharing \subs{federated learning}, Model/Data Exchange & {\nTransp} & Help auditors interact with and analyze proprietary information about the data or model with centralized infrastructure. & NIST's Face Recognition Vendor Test \citep{ngan_ongoing_2020}, Google AI Test Kitchen \citep{warkentin_join_2022}, Airbnb's Project Lighthouse \citep{airbnb_new_2020} \\
\hline

Data Collection & Field Data Collection \subs{scraping, donation, interviews/surveys, compelled disclosure}, Bot Deployment, Simulation & {\nData} & Help auditors collect information about a model’s interactions with its subjects. & Mozilla's YouTube Regrets \citep{mozilla_foundation_youtube_2021}, \href{https://tracking.exposed/}{Tracking Exposed}, \href{https://www.selenium.dev/}{Selenium}, Meta's Web-Enabled Simulation \citep{ahlgren_wes_2020} \\ 
\hline

Performance Analysis & Accuracy Evaluation \subs{A/B testing, benchmarks, adversarial testing, monitoring}, Explainability \subs{models, training data}, Fairness, Qualitative Analysis & {\nPerf} & Help auditors evaluate and explain model behavior through the calculation of performance metrics. & \href{https://wandb.ai}{Weights \& Biases}, Meta's \href{https://dynabench.org/}{DynaBench}, \href{https://foolbox.jonasrauber.de/}{Foolbox}, \href{https://fairlearn.org/}{Fairlearn}, IBM's \href{https://github.com/Trusted-AI/AIF360}{AI Fairness 360}, Hugging Face's ROOTS \citep{piktus_roots_2023}, Google PAIR's \href{https://pair-code.github.io/lit/}{Language Interpretability Tool}\\ 
\hline

Audit Communication & Dataset Visualization, Audit Reporting & {\nComm} & Help auditors communicate the results of an audit to a broader audience. & Google PAIR's \href{https://pair-code.github.io/facets/}{FACETS} \\
\hline

Advocacy & Organizing/Resistance, Community Spaces, Legal Search & {\nAdvocacy} & Help organize community action and other accountability measures in response to discovered harms. & \href{https://gigbox.media.mit.edu/app/}{Gigbox}, \href{https://www.withpara.com/}{Para}, \href{https://adnauseam.io/}{Adnauseam}, \href{https://www.btah.org/}{Benefits Tech Advocacy Hub} \\

\bottomrule
\end{tabularx}
\end{table*}

Second, while we found many freely available and open source tools ({\osPropTotal} of our dataset), auditors highlighted the messy, context-specific nature of their actual audit tool use: \inlinequote{Many approaches were not necessarily principled. They were quite ad hoc}{20}. Even when open source tools existed, auditors often preferred to build their own tooling solutions: \inlinequote{%
if we tried to use the existing stuff, it would just complicate that process}{25}.
For some, existing tools were inadequate for the complexity and scale of the systems being evaluated: \inlinequote{Most often we try to use open source tools, but that's very different than a data pipeline that... curates data on millions of [users] every day}{3}. 
In each stage of the audit process, auditors encountered practical challenges and development gaps between their needs and the landscape of available tools.

In the remainder of this section, we detail the challenges practitioners faced in each stage, compare their experiences to the existing landscape of AI audit tools, and discuss the implications of our findings for the practice and study of AI auditing.

\begin{figure*}[!t]
  \centering
  \includegraphics[width=\linewidth]{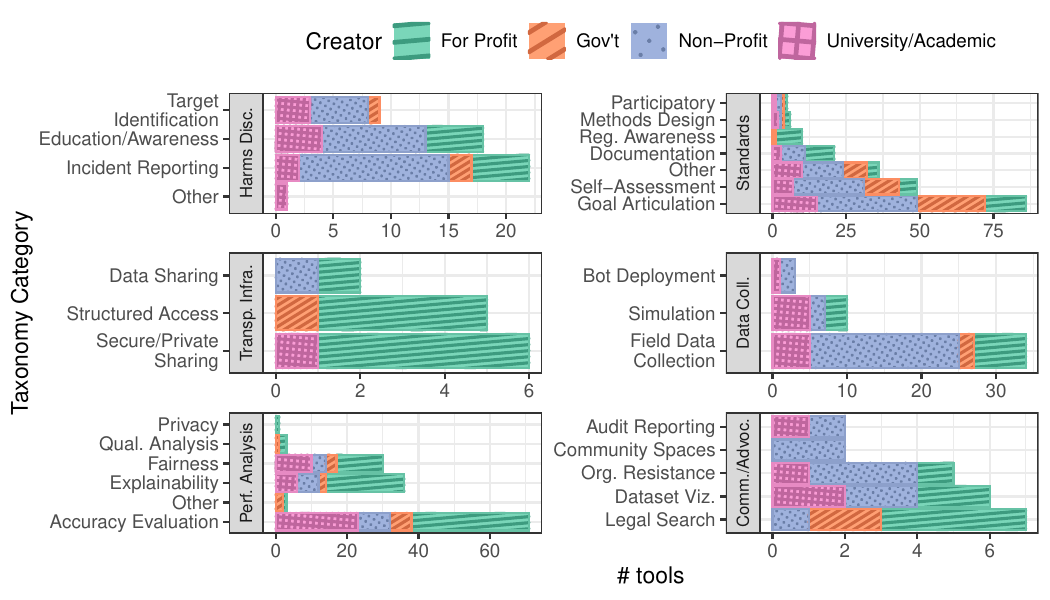}
  \caption{Number of tools in each category within each stage of our taxonomy, grouped by type of organization. \taxonomyDisclaimer{} Note that the scales differ---the Standards and Performance Analysis stages contain many more tools than the others. Nonprofit and university/academic developers account for relatively more Harms Discovery and Data Collection tools. %
  For-profit developers contribute %
  relatively more Performance Analysis and Transparency Infrastructure tools.}
  \Description{In the provided bar chart, we display the number of tools categorized under six major taxonomy categories that includes, 'Harm Discovery', 'Transparency Infrastructure', 'Performance Analysis', 'Standards Identification', 'Data Collection' and 'Audit Communication'. These categories are further divided based on their organization type, which can belong to one of four groups: 'For profit,' 'Government,' 'Non Profit,' or 'University/Academic.' The chart visually represents the quantity of tools in each of these categories using vertical  bars, allowing readers to understand the distribution of tools across different taxonomies and organizational types.}
  \label{fig:profit-count-second}
\end{figure*}

\subsection{Harms Discovery}
\label{sec:harms}
Auditing an AI system first requires identifying %
the %
system that should be subject to scrutiny and identifying its potential harms. This task can be especially difficult for \emph{external} auditors who may not know where AI systems are in use or what their impacts might be. Tools for Harms Discovery ($N=\nHarms$) help identify and select targets for audits and support the identification, characterization, and prioritization of potential harms to investigate. This category includes tools for Education \& Awareness (to engage affected stakeholders in articulating harms), Incident Reporting (to gather reports of algorithmic harms from users and the public, e.g., through bug bounties or incident databases \citep{charlie_pownall_ai_2021}), and Target Identification (e.g., the Algorithm Tips database contains a list of deployed systems in the U.S.).
Compared to other stages of our taxonomy (Fig.~\ref{fig:profit-count-second}), nonprofits contributed %
significantly to creating and maintaining these types of tools in our dataset ({\notprofitPropHarms} not-for-profit; see Fig.~\ref{fig:profit-employees}).

\inlinesection{Facilitating more participatory audits.}
Auditors recognized that to comprehensively identify AI-related harms, they must engage with those directly impacted. %
Participation in harms discovery had two main benefits for auditors. First, participation helped auditors anticipate a broader range of possible harms: \inlinequote{Different types of biases are going to manifest, and accordingly it requires ... diverse groups from society, to understand their experiences and expectations in these settings and how they can be impacted}{20}.
Second, participation helped make audits more ``context-dependent'' and inclusive by providing thorough understanding of how an AI system interacts with impacted groups.
Participation %
also helps instill confidence in the audit process %
and foster engagement with subsequent accountability efforts.
Our tool survey %
surfaced %
some tools for participatory incident reporting. This includes %
the bug bounty platform HackerOne, which Twitter used to crowd audit its image cropping algorithm \citep{twitter_twitter_2021}, 
and the AI Incident Database, a collection of reports of AI harms. However, we found fewer tools designed specifically for identifying and collaborating with affected users, such as the American Civil Liberties Union (ACLU) of Washington's Algorithmic Equity Toolkit \citep{barghouti_algorithmic_2020}.

\inlinesection{Avoiding participation washing.} %
Participants also recognized the challenge of designing methodologies and tools without exploitation, tokenization, or other forms of ``participation washing'' \citep{sloane_participation_2022}, highlighting the need for fair compensation and participatory algorithmic development, in addition to participatory auditing. One participant emphasized the importance of \inlinequote{an iterative process of doing longer-term, ongoing auditing or observation of algorithmic behavior, and then using that to feed into tweaks, or changes, or even big shifts in where and how [audit] outcomes are used}{17}.

\inlinesection{Implications.}
Limited access to information about AI systems poses a fundamental barrier to conducting comprehensive audits.
Without a clear understanding of which AI systems require auditing, there is a risk of overlooking critical systems that have significant societal implications. While we did find multiple popular databases for recording and collating incidents of harm (such as the AI Incident Database), these databases record harms after they have already occurred, often rely on second-hand reports, and may not delve deeply into causes of harms or impacts \citep{turri_why_2023}.\footnote{Note: The creators of the AI and Algorithmic Incident and Controversies Repository dispute the characterization of their tool in \citet{turri_why_2023}.}

Several participants proposed mandating that corporations disclose AI system use, including information about model versions, anticipated use cases, expected number of users, and past audit results. Researchers and policymakers may also explore mechanisms for centralized, proactive documentation and mandatory, standardized incident reporting for both private firms and government agencies \citep{turri_why_2023}. This %
ensures that current federal AI transparency requirements are actually implemented \citep{lawrence_bureaucratic_2023}. 
Additionally, leveraging mechanisms such as Freedom of Information Act (FOIA) requests could facilitate access to information held by public institutions or government agencies regarding the use of AI systems.
Future work could also develop and study systems for fair, inclusive community participation in auditing, the path most often suggested by practitioners for identifying systems and their harms.

\subsection{Standards Identification \& Management}
\label{Standards}
Auditors also used tools to formulate principles and norms to guide their investigations. While HCI research has not traditionally included frameworks and guidelines as tools for AI, Standards Identification \& Management was a key focus for participants and comprised the largest collection of tools in our dataset.
Standards Identification \& Management ($N=\nStandards$), includes tools for Goal Articulation (e.g., broad principles statements), Self-Assessment (more specific procedural assessment tools, such as Microsoft's AI Fairness Checklist \citep{madaio_co-designing_2020}), Documentation (e.g., Model Cards \citep{mitchell_model_2019}), Regulatory Awareness (tools, often paid services, for discovering and monitoring relevant regulations), Methods Design (standards for audit methodology), and Participatory Standard-Setting (methods for developing standards in collaboration with affected groups, such as Microsoft's Community Jury \citep{cass_community_2022}).

This category includes both internal organization standards and principles and formal national or international standards. Tools such as NIST’s Risk Management Framework (RMF) and relevant ISO standards are referenced extensively by auditing organizations \citep{babl_auditor_certification}, and regulators in Singapore and other nations have invested in similar tools as part of compliance and oversight frameworks \citep{infocomm_media_development_authority_cataloguing_2023, kaye_risky_2023}. While the weight and enforceability of these standards differ, they are united by a shared goal of defining audit methodologies and expectations for system performance.

\inlinesection{{Need for more context-specific standards.}}
Despite the large number of standards and evaluation frameworks surfaced in our tool survey, auditors still felt that evaluation frameworks needed refinement. Auditors emphasized the importance of standardized evaluation frameworks that provide clarity and consistency: \inlinequote{I think standardization is a big [concern]...}{23} Many wished to streamline the auditing process by offering predefined structures and templates for assessment, which are essential for conducting audits effectively and facilitating communication.
Most commonly found were goal-articulating ``principle statements'' ($N=\nStandardsGoal$), self-assessment checklists ($N=\nStandardsAssessment$), and similar documents,
while methods for participatory standard-setting ($N=\nStandardsParticipatory$) were comparatively rare in our dataset.
The Standards Identification tools we found were particularly general in their applications, compared to other categories of tools---the principle statements, checklists, and similar resources we found were usually developed without a specific kind of target system in mind (Fig.~\ref{fig:target-count-second}). Some participants found these tools too broad to easily apply:
\blockquote{%
    I think what I've seen is that companies and institutions... really, really struggle to understand, `What should we even do when it comes to auditing or evaluating the use of machine learning in our organizations?' And while a template or a checklist is not the right answer, a lot of them don't even know where to start...
    And so, (it helps) when you have a tool that...has some built in frameworks.
}{8}

\inlinesection{{Need for more standards beyond fairness.}}
Some participants also thought that assessments templates and checklists were focused too narrowly on fairness assessment. One civil society auditor said, \inlinequote{I would love some guidance on audits generally and tools that describe the non-fairness components of an audit...}{6} Another wished for %
resources that covered criteria such as explainability, privacy, and transparency: \inlinequote{Currently we have to.. find open source tools and put them together ourselves, and you need to have expertise to know what to look for}{21}.

\inlinesection{{Need for clear and consistent regulatory guidance.}}
Despite some participants' desire for official frameworks like NIST's Risk Management Framework (RMF) \citep{tabassi_ai_2023}, multiple participants commented on the difficulty of harmonizing current or expected regulatory guidance with practical implementation. Regulatory guidance itself may function as a tool, providing a framework for compliance with emerging policies. One civil society auditor asked:
\blockquote{%
    What are we evaluating for? And the question of, even when we have some kind of legal or other benchmark in mind, what are the metrics, and what are the benchmarks and other ways in which to evaluate, technical and otherwise, which also remain quite unclear? We're seeing the ready adoption of audit language into policies, so that just kind of makes us nervous... What are we auditing for?}
{2}
Several emphasized the necessity of regulatory entities being more forthcoming in defining best practices. Some sought the guidance of the regulatory bodies in the domains they operate to establish their own frameworks but often struggle to translate industry expectations into meaningful standards for evaluation. Multiple participants felt there was a \inlinequote{culture or communications gap between the legal and compliance people on one side and the engineers on the other}{12}. As a result, some hoped for more collaborative approaches instead of command-and-control prescriptions: \inlinequote{%
    Half the time we reach out [to regulators], and there's just no one to contact... it's just a black hole...
}{13}.

\inlinesection{Implications.} Standards for evaluating AI systems must be simultaneously holistic, context-specific, inclusive, and compatible with practice. Some of the tools we found made advances in one or more of these dimensions, but few accomplished all three. Microsoft's AI Fairness checklist, for example, was co-designed with practitioners \citep{madaio_co-designing_2020} in an effort to be more compatible with practical challenges, but many of the other standards frameworks we found did not obviously consult practitioners and fewer involved affected stakeholders. Likewise, while NIST's AI RMF includes safety, security, reliability, transparency, explainability, and privacy, in addition to fairness, its guidance for specific evaluation techniques remains fairly broad \citep{tabassi_ai_2023}.

Research could continue to explore how regulatory standards could be translated into concrete metrics and other effective guidance for industry \citep{wachter_bias_2021, guha_ai_2023}. One participant at a large tech company, for example, preferred a \inlinequote{must, could, should}{14} structure for regulatory guidance: a non-technical legal minimum (``must'') accompanied by more precise technical paths to compliance (``should'') and a set of ideal best practices for high performers and innovators (``could'').

\subsection{Data Collection \& Transparency Infrastructure}
\label{Data}
Gathering empirical evidence is a key step in AI auditing, but often poses the most significant challenge in practice. %
When model operators were unwilling or unable to release relevant documentation and other evidence, auditors turned to two main classes of tools to help.

Tools for Transparency Infrastructure ($N=\nTransp$) are interfaces and databases hosted by model operators that allow controlled access to relevant data.
This category includes tools for Structured or Application Programming Interface (API) Access (tools that allow auditors to interact with models and live systems, such as Google's AI Test Kitchen \citep{warkentin_join_2022}), tools for Data Sharing (platforms or trusts for hosting models and related data, such as the Gig Economy Data Hub \citep{noauthor_gig_2021}), and tools for Secure \& Private sharing (tools that help mitigate concerns with sharing data, such as Airbnb's Project Lighthouse \citep{airbnb_new_2020}).

More commonly, though, tools for Data Collection ($N=\nData$), helped \emph{external} auditors in particular gather information \emph{outside} auditee-controlled interfaces. %
These tools help auditors gather data about model behavior, including relevant information not routinely collected by model operators. This category includes tools for Field Data Collection, which collect data from real systems and real users---including tools for Data Donation (such as Mozilla's YouTube Regrets project \citep{mozilla_foundation_youtube_2021}), Data Scraping (e.g., Tracking Exposed \citep{agosti_tracking_2023}), Interviews/Surveys, and Compelled Transparency (e.g. tools such as MuckRock, which facilitates public records requests). We also found tools for Simulation (e.g. Meta's Web Enabled Simulation platform for simulating interactions on Facebook \citep{ahlgren_wes_2020}) and Bot Deployment (tools used for sock puppet auditing \citep{bandy_problematic_2021}, such as Selenium or Appium), both used to test systems with artificial or semi-artificial interactions.

\inlinesection{{Need for uncompromised data access.}}
Data collection tools aim to address one of the challenges most frequently mentioned by our participants: the difficulty of accessing data and other vital information required to conduct meaningfully independent audits. Transparency Infrastructure tools provide external auditors with controlled access to models and data---especially for online platforms, in our dataset (Fig.~\ref{fig:target-count-second})---but they require investment from model operators. Despite court orders and regulations like Article 40 of the Digital Service Act~\citep{leerssen_digital_2023} that require the construction of transparency tools, one participant noted that key APIs used for auditing are becoming more costly and undependable:
\blockquote{There's a direct impact on shutting off the access to information that affects people doing audits... we saw that with Reddit charging for their API and shutting down... Twitter charging astronomical, now, amounts for their API. It's because everybody is scraping public Reddit and public Twitter to train large [AI] models...}{24}
They wished for \inlinequote{access to platform data... a context under which people can do controlled experiments, using data that is provided by platforms directly}{24}.

Auditors also said that corporate control over APIs undermined their independence in conducting audits which aligns  with our definition of an audit (Def.~\ref{def:audit}). A civil society auditor said,
\blockquote{I wish I had something to force people to give me their data... Part of the problem of auditing in general is that the only people who get let in are usually the people who are willing to say nice things about whatever the technology is being audited.}{6}
In practice, participants reported that key details, such as data sampling methods, data provenance, model versioning, metrics, and design justifications, were often omitted or only partially disclosed.
And currently, the vetting process for API use often requires the auditor to disclose their intent for the evaluation in advance, which may compromise the integrity of the study.
\blockquote{I'm very much concerned that what's going to happen is.. [platforms have] given all this access, and there will actually be... more of a cover up than there is now... They have so much power in this conversation to just share whatever information they want.}{7}

Instead, one participant's ideal was an \inlinequote{inspectability API}{7}, a required, standardized interface to allow researchers to interact with online platforms, including the ability to test different profiles, geographies, and other variables needed to evaluate disparate treatment, misinformation, and other algorithmic harms.
Similarly, an auditor at a startup wished for centralized data archive available to the public: \blockquote{%
    If I really had to paint my perfect vision, it would be an independent database archive, or whatever, of all the relevant data. And then... different parts of society can tap into it... I think what you want is a lot of different innovative organizations and people and builders taking this data and building useful things with it, rather than a single one.
}{12}

\begin{figure}[!t]
  \centering
  \begin{subfigure}{0.48\textwidth}
      \centering
      \includegraphics[width=\linewidth]{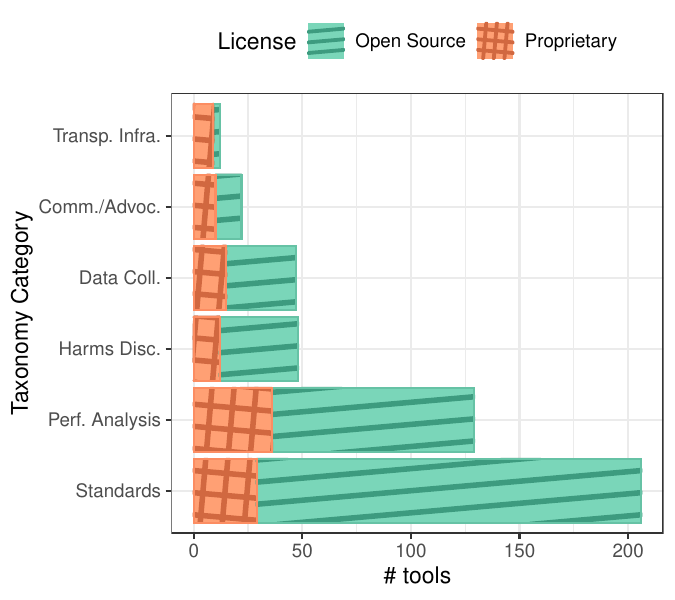}\\
      \caption{Number of open source tools in each taxonomy stage. \taxonomyDisclaimer{}}
      \Description{Bar graph showing the number of tools in each category based on the license types, 'Open Source' and 'Proprietary'.}
  \end{subfigure}
  \begin{subfigure}{0.48\textwidth}
      \centering
      \includegraphics[width=\linewidth]{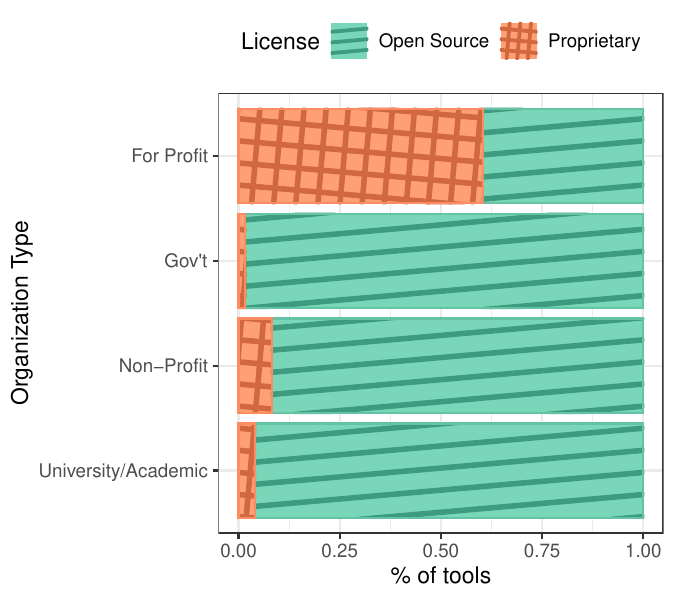}\\
      \Description{Bar graph showing the percentage of tool based on their license type, 'Open Source' and 'Proprietary'. Stratified by organization type.}
      \caption{Percentage of open source tools by organization type.}
  \end{subfigure}
  \caption{Tool licensing by taxonomy stage (top) and by organization type (bottom).}
  \label{fig:profit-license}
\end{figure}

\inlinesection{Challenges with independent data collection.}
Rather than rely on the model operator to provide access, auditors---especially \emph{external} auditors---often turned to tools for Data Collection to obtain evidence themselves, sometimes developing and sharing their own tools and processes.
Unlike the tools we found for Transparency Infrastructure, which were mostly not open source ({\osPropTransp} open source in our dataset; see Fig.~\ref{fig:profit-license}), tools for Data Collection were much more likely to be available under an open source license ({\osPropData} open source in our dataset). Tools for Data Collection---most not built specifically for AI auditing---also comprised the most popular Github repositories in our dataset (Table~\ref{tab:top-repos}). For example, auditors used Selenium \citep{noauthor_selenium_2023}, a popular collection of open source tools for browser automation, to simulate user profiles while scraping data (known as a ``sock puppet'' audit \citep{bandy_problematic_2021}).

While this approach gave auditors more freedom, it could also take more effort. Some of the tools we found---such as the Markup's Citizen Browser \citep{the_markup_citizen_2022}, a data donation platform---were built from scratch to collect specific kinds of data for auditing. Existing tools for data scraping were helpful but often required extensive adaptation:
\blockquote{%
     We almost always have to build custom scrapers to collect data... There's some templates right out there for these scrapers, and then you usually have to customize them. And then there's a huge amount of work to keep them alive... They break all the time with all these edge cases. And so they're really a pain. I don't really know that there's a way to solve that.
}{7}
Despite these difficulties, external auditors---particularly the journalists in our sample---still saw advantages in independent data collection. The same journalist noted: \blockquote{%
    There's a lot of requests for inside access, as if that's the only way to do this type of thing, but in reality inside access can actually be a trap... your view inside the room is actually really limited. And so I have come to believe... that actually doing analysis from the outside can often be way more revealing... I usually never have insight into the algorithm itself, but I can do analysis on the outputs. And to me, that's the right place for a journalist to operate, because the outputs are the real-life impact.
}{7}

\inlinesection{{Challenges with understanding and processing data.}}
Even when they had access to the data they wanted, participants noted that basic challenges involved with managing and analyzing data required more labor than any other task. A government auditor said, \inlinequote{90\% of the work is figuring out what different tables are, and what different columns are, and working out whether it makes sense to join certain things. And then figuring out some meaningful metrics that we can draw from that data...}{25}.
Participants also spent lots of time reviewing and requesting additional documents from audit targets.

Multiple participants wished for tools to help with tasks
such as data collection and cleaning: \inlinequote{most of the value of data infrastructure is literally cleaning data}{3}.
One auditor hoped for innovation in data quality management: \inlinequote{There are custom scrapers, a lot of human data quality work, and one thing that I have really wanted to do and never been able to do is try to figure out ways to get that data quality work done in more interesting ways}{7}. 

Data quality concerns intersected with concerns about audit integrity. \inlinequote{I think we often have to worry about... [whether] what we see is what we think it is}{15}.
With recent datasets scaling up to staggering sizes, this concern has become more acute and auditors commented on how manual analysis was no longer feasible: \inlinequote{Given how much data we are able to process, we need new methods to analyze the data curation process and what kind of problems data comes with. And then we need tools to detect what is synthetic, what is real}{20}.

\inlinesection{Risk of retaliation.} Some external auditors also worried that external data collection tools---particularly tools for data scraping or data donation---may violate terms of service set by platforms and result in legal liability or retaliation. 
Auditors expressed concerns about legal risks from existing laws and regulations, particularly the Computer Fraud and Abuse Act (CFAA) %
and other privacy laws. Under the CFAA, for example, an auditor who violates a platforms' terms of service may be held criminally liable. 

It is difficult for auditors to determine what methods---such as public data scraping or even data donation---may be deemed ``unauthorized'' and criminal under the CFAA unless the audited organization grants authorization. In \emph{Sandvig v. Barr} (2019), for example, the ACLU sued to allow researchers to set up false accounts (sock puppets) to audit computer algorithms \citep{noauthor_sandvig_2020, american_civil_liberties_union_sandvig_2019}.
In 2021, Facebook used exactly this term of art---``unauthorized''---several times in its justification for disabling the accounts of a group of researchers auditing its advertising algorithms with a data donation tool called Ad Observer \citep{clark_research_2021, edelson_we_2021}.
Facebook initially insinuated that disabling the researchers' accounts was required by an Federal Trade Commission (FTC) consent decree, backtracking only after the FTC called the statement inaccurate \citep{levine_letter_2021}.
One civil society auditor said, \blockquote{Even if the legal risks have not been acted upon as much, there's cases everyone points to in terms of attacks against researchers. It's a matter of time before it ramps up. As soon as our work becomes threatening enough, that's when it all really starts.}{24}

As a result, auditors engaged in external data collection had to take great steps to guard against liability and retaliation. One journalist said, \inlinequote{[The CFAA] is an incredible legal hangover for the type of work I do... how much lawyering I need to even get one tool off the ground is insane}{7}. Auditors also expressed hesitation about reforms that give platforms more control over what data is released. The same journalist continued,
\blockquote{%
    Exemption from [the CFAA] would honestly be more helpful than these platform access roles that the E.U. is claiming that they're going to offer [e.g., in the Digital Services Act], which I am very skeptical about... I just feel like the history of these things is that when platforms have been required to provide API access, they have somehow always made it impossible to do real accountability.
}{7}
Even auditors hired internally may assume some degree of personal risk. One civil society auditor noted, for example: \blockquote{%
    Another really frequent sort of question that I get [from organizations]... is what is my liability around doing this kind of [audit work]? And frankly, to your earlier question about building in-house versus contracting, that's another main [reason to contract]... it's like, okay, we're still going to do this thing, but just sort of outsource it. So I think that just remains like a really open question that people doing this kind of work are carrying a lot of legal risk in doing so.
}{24}

\inlinesection{Implications.} Future research could explore tools and processes for not only facilitating access to data---especially independently, through scraping or simulation---but also for ensuring data quality and integrity. Participants specifically wished for more tooling for data donation and user-driven auditing, a nascent area of research in human-computer interaction \citep{lam_sociotechnical_2023, devos_toward_2022, deng_understanding_2023}.
Auditors also faced challenges common to data work in general, and research on practices surrounding data quality and data integrity may be applied specifically to discrimination testing and AI auditing methods.

Other challenges were more particular to auditing work. Future work could explore how auditors request information and interact with model operators. Transparency Infrastructure in particular is a nascent area of tooling that may become more common in auditing practice as AI regulation develops and as barriers to external data collection mount. Independent research may help guide these tools into more trustworthy mechanisms for disclosure, even as policymakers can ensure platform-controlled tools are not the only avenue for scrutiny.

\subsection{Performance Analysis}
\label{Performance}
Tools for Performance Analysis ($N=\nPerf$) are designed to help auditors evaluate and explain model behavior, usually through the calculation of quantiative metrics related to accuracy/safety ($N=\nPerfAcc$), explainability ($N=\nPerfExp$), or fairness ($N=\nPerfFair$). This category includes tools for Fairness Evaluation, Accuracy Evaluation (including tools for A/B testing, benchmarking, and model monitoring, such as Meta's Dynabench or the Linux Foundation's Adversarial Robustness Toolbox), Explainability (tools for explaining the behavior of a model, such as IBM's AI Explainability 360, or for exploring training data, such as Hugging Face's ROOTS search tool \citep{piktus_roots_2023}), and Qualitative Analysis.

\inlinesection{{Concerns about methodological integrity.}} Despite the many tools developed for Performance Analysis---including the most popular AI-specific Github repositories in our dataset (e.g., OpenAI Evals \citep{openai_gpt-4_2023}; see Table~\ref{tab:top-repos})---practitioners expressed a need for more robust, well-vetted tools and methodologies.
Internal auditors in particular had concerns about the validity, reproducibility,  transparency, and trustworthiness of the methods used in popular Performance Analysis tools. One auditor at a tech startup said, \inlinequote{I'm still not convinced of the validity, even, of some of those methods}{4} used in tools for monitoring and validation.

For example, the most popular Performance Analysis tool we found on Github is SHAP (SHapley Additive exPlanations), a game-theoretic method for measuring feature importance in a model \citep{lundberg_unified_2017} (see Table~\ref{tab:top-repos}). But as \citet{kumar_problems_2020} argue, Shapley values are prone to misuse and may be unsuitable for normative evaluation.
Interpretability and explainability methods promoted by popular tools vary widely in their goals \citep{lipton_mythos_2018} and, like many ``snake oil'' AI products \citep{kaltheuner_fake_2021, narayanan_ai_2024, stark_physiognomic_2021}, may encourage false confidence in their users \citep{ghassemi_false_2021}. Yet explainability tools such as SHAP are often suggested in official regulatory guidance \citep{kaye_risky_2023}.

Some participants put methodological deficiencies down to differences in the rigor employed by the various disciplines involved with auditing. An auditor in civil society said, \inlinequote{The bar of the kind of threshold of... validity of findings and novelty of findings is much higher in academia than it is for civil society}{24}.
Participants had concerns about maintenance, effectiveness of automated monitoring processes, and the efficacy of synthetic data for representing real users instead of functioning as \inlinequote{an academic exercise}{4}: \inlinequote{You can perturb all the different inputs you want. But they might not be realistic combinations of features for people who are actually using the system}{25}. 

\inlinesection{Need for inspectable, reproducible methods.} To allay methodological concerns, several participants emphasized the importance of open-sourcing tools for others in the community to inspect. Some of our industry participants had reproducibility in mind when designing evaluation procedures: \inlinequote{You want to iterate... but you know that also makes the results less reproducible. And are you being deceptive then, if you [refer in published evaluations] to a model that's different from the one that people analyze?}{5} However, the Performance Analysis tools we found ({\profitPropPerf} of which were built by for-profit organizations) were disproportionately \emph{not} open source 
compared to other tools in our dataset,
especially tools for Explainability (Fig.~\ref{fig:license-count-second}).

\inlinesection{Need for more analysis tools beyond fairness \& explainability.} Similar to Standards Identification \& Management, our participants wished for tools to help evaluate a broader spectrum of criteria for AI systems.
The Performance Analysis tools we found---the most popular of which were built and maintained by disproportionately large,
for-profit firms (Figures~\ref{fig:profit-employees})---often focused on a narrow set of technical fairness definitions and explainability methods popularized in academic literature. The tools we surveyed in this category often overlooked entire other areas of concern, including the basic functionality of the model \citep{raji_fallacy_2022} as well as other methods of evaluation. For example, tools specifically devoted to qualitative---as opposed to quantitative---analysis were much harder to find ($N=\nPerfQual$) and rarely mentioned by our participants.

\inlinesection{Implications.} While there are multiple studies on the use of tools for fairness evaluation \citep{holstein_improving_2019, lee_landscape_2021, deng_exploring_2022} and explainability \citep{bertrand_selective_2023, wang_designing_2019, liao_questioning_2020, kaur_interpreting_2020, smith-renner_no_2020, kim_help_2023}, fewer studies examine how practitioners evaluate other criteria such as basic functionality \citep{raji_fallacy_2022}, safety, privacy, or recourse, just as fewer tools exist for this purpose. Future work could develop and investigate tools for a broader range of evaluation criteria.
Future work could also explore practitioners' standards for audit tooling \citep{kaye_risky_2023}, and policymakers may develop standards that require academic peer review or vetting by regulatory bodies for audit tooling.

Moreover, research must examine further how tools may contribute to
``audit washing,'' the use of auditing procedures to legitimize unethical practices \citep{goodman_ai_2022}. 
A tool for accuracy evaluation, for example, may be used to analyze the accuracy of dubious technology for predicting ``criminality'' or ``trustworthiness'' without questioning underlying ethical issues with these applications \citep{stark_physiognomic_2021, wang_against_2023}. In general, tools may claim to provide auditing capabilities---using terms such as fairness, safety, or explainability---while failing to conduct evaluations that meaningfully contend with power dynamics and institutional barriers to accountability \citep{wong_seeing_2023}.

\subsection{Audit Communication \& Advocacy}
\label{sec:audit}
Some of the most crucial accountability work of an AI audit comes after empirical evaluation is complete. We found two emerging sets of tools that begin to address this important stage of AI auditing: tools for Audit Communication, to effectively translate audit results to a broader audience, and tools for Advocacy, for reporting and campaigning for consequential outcomes in response to audit results. Tools to facilitate Audit Communication were the rarest in our dataset ($N=\nComm$), and consist mostly of tools for Dataset Visualization (e.g. Google's FACETS \citep{noauthor_facets_2023}%
) and 
Audit Reporting (e.g., the ACLU's repository of NYC Local Law 144 hiring bias audit reports \citep{gerchick_tracking_2024}). We found more tools for Advocacy ($N=\nAdvocacy$)---including Community Spaces (e.g., the Benefits Tech Advocacy Hub \citep{noauthor_benefits_2023}, which facilitates collaboration between advocates who oppose algorithm-based cuts to public benefits), tools for Organizing/Resistance (e.g., the Algorithmic Ecology framework \citep{stop_lapd_spying_coalition_algorithmic_2020}, a tool for mapping the non-technical dimensions of algorithmic impact),
and tools for Legal Search (e.g., the generic case database Westlaw often used to identify relevant precedent for legal redress)---but still fewer than we found in other stages of our taxonomy.

These types of tools were also mentioned less often in our interviews, compared to preceding stages of our taxonomy. Still, auditors wanted their evaluation work to inform consequential judgments, in line with our definition of an audit (Def.~\ref{def:audit}). As one put it, \inlinequote{in the business of designing audits, it should be as important to design the consequences and penalties that accompany these audits}{2}.
For the auditors we spoke to, tooling in this stage of auditing was mostly aspirational.

\inlinesection{Tools and resources for community building.} Auditors especially wished for resources that would bring together the diverse, interdisciplinary groups involved in auditing, similar to the few tools for Community Spaces we found in our tool survey. As one auditor put it, \inlinequote{We have to have people in the accountability business}{7}.
Auditors hoped greater communication could help unite the profession around policy developments, shared language, standards, and goals that could improve the impact of their work.
One auditor at a tech startup said, \inlinequote{[NIST] has AI guidelines that come out, and we work with them... we send in comments, we give talks, all that kind of stuff. I think that's an important part of the auditing community}{4}. Another described a workshop attended by civil society, academics, and consulting firms to help prepare for legislation in the European Union (P23). One auditor hoped that communication could lead to shared tooling: \inlinequote{How do we bring these interdisciplinary communities together so that we can use tools together?}{20}

\inlinesection{Implications.} Communicating audit findings, lessons, and insights learned can help build trust and validate audit findings, recommendations, and subsequent interventions. Audit report repositories could expand the forum holding model operators accountable, allowing policymakers, journalists, and other public stakeholders to engage with evaluations more easily. 
Embracing public evaluation results also helps
audit practitioners to learn from each other’s experiences.
Despite these benefits, tooling to support these stages is rare. While we found some domain-specific spaces where auditors can interact---the Benefits Tech Advocacy Hub \citep{noauthor_benefits_2023}, for example---we found few audit reporting tools or tools for facilitating communication between auditors, journalists, and activists. Future design work and research could explore these emergent categories. Academic research could also explore in more detail the specific mechanisms of audit communication that are most likely to result in meaningful change---such as including concrete demands for action~\citep{raji_actionable_2022}---and imagine new tools to support those mechanisms.

\section{Discussion}

The HCI community has historically contributed key research to the design and development of \emph{AI audit tools} and helped define the concept of AI auditing~\cite{sandvig_auditing_2014, lee_landscape_2021,holstein_improving_2019, wong_seeing_2023}. In this section, we specifically discuss the important takeaways for that community %
in particular, as well as broader lessons for other stakeholders, including policymakers, audit practitioners, and funders. %

\subsection{Moving beyond evaluation, towards accountability}

\citet{costanza-chock_who_2022} found that over 65\% of surveyed AI audit practitioners felt that ``accountability'' (defined as a  ``commitment from auditee to address problems covered by audit within set time'') was a top unmet need in their AI auditing work. This echoes a theme  repeated several times throughout our interviews---AI auditors care deeply about accountability but struggled to achieve it. %

Despite searching deliberately for non-evaluation tools, we found more than \emph{five times} as many tools in the evaluation stages of the AI audit process as we did tools for harms discovery, audit communication, or advocacy.
Perhaps unsurprisingly, these are also the stages of the audit process that participants described as most requiring contextual awareness and typically under-studied participatory and community engagement methods.
Research and development related to these and other practical challenges could bolster practitioners' accountability efforts. Promising new directions for HCI research and policy include:

\begin{itemize}

\item \inlinesection{Studying and developing tools for harms discovery, audit communication, and advocacy.}
Our tool survey identifies several neglected categories of tools---particularly tools for Incident Reporting, Education/Awareness, Target Identification, and Audit Communication---that are worthy subjects for future HCI research. For instance, promising recent research explores the existing limitations~\citep{turri_why_2023} and educational applications~\citep{feffer_ai_2023} of incident reporting databases, but little work explores
complementary technical infrastructure, such as, for example, the AI inventories often used by journalists (such as Algorithm Tips) and previously required for federal agencies~\citep{biden_executive_2023}.
Likewise, auditors envisioned audit report databases as accountability tools to facilitate the amalgamation and communication of audit findings to key stakeholders, mirroring interventions such as the U.S. Security and Exchange Commission (SEC) EDGAR database for financial accounting audits ~\citep{raji_outsider_2022}. These gaps in development also present meaningful opportunities for further investment and institutionalization by policy-makers and funders.

\item \inlinesection{Validating existing tools in practice.}
Audits have limited impact if their results are not reliable or meaningfully connected to real world requirements \citep{raji_actionable_2022}. Unreliable performance or accuracy analysis tools that fail to meaningfully assess the audit target operate as misleading ``rubber stamps'' for vendors and lead to ``audit washing''~\cite{goodman_ai_2022}, posing a %
serious challenge to the legitimacy of audit results.
There is a growing opportunity for HCI researchers to explore ways that AI audit practices interact with existing accountability processes such as
litigation or regulatory compliance. For instance, several tools surfaced in our survey (e.g., from Holistic.ai, Credo.ai) were explicitly marketed for use for NYC Local Law 144 compliance, but studies of audit practice suggest the produced measures may not be reliable or legally compatible~\citep{xiang_legal_2019, groves_auditing_2024, wright_null_2024}. Researchers might also investigate the validity and effectiveness of government-sponsored tooling ~\citep{kaye_risky_2023} and court-mandated transparency infrastructure (e.g., Facebook Ad Library, built after settlements with civil rights groups \citep{sandberg_doing_2019}).

\item \inlinesection{Developing participatory methods for audit work.}

Given the broader calls for a participatory turn in AI development ~\cite{delgado_participatory_2023, kulynych_participatory_2020, birhane_power_2022}, it is no surprise that participation is an increasing focus for AI audit practitioners and audit tool developers as well. Recent HCI work on user auditing \citep{deng_understanding_2023}, such as the ``WeAudit'' tool ~\cite{devos_toward_2022}, exemplifies this shift and demonstrates the possibility of designing for a more participatory AI audit process~\cite{shen_everyday_2021,shen_publics---loop_2022,deng_understanding_2023}. 
Policymakers can also further emphasize participation as a requirement in audit guidance and invest in tools that support participatory methods.

\item \inlinesection{Open \& reproducible practices for AI audit tools.}
Some participants were concerned about the efficacy of many AI audit tools---particularly tools whose methods were not made available for public scrutiny. (Tools we found for Performance Analysis were less likely to be open source; Fig.~\ref{fig:profit-license}). Making AI audit tools publicly available enables both external collaboration and third-party validation. Open tooling practices may also contribute to knowledge-sharing, standards-setting, transparency, accessibility, and trust, but can have complex interactions with power and oversight that are worthy of further study \citep{widder_why_2024}.
Researchers, policymakers, foundations, and other stakeholders developing audit tools should prioritize open practices.
Policymakers could also consider requiring that published audit reports include
clear explanations of auditors’ methods and tools.

\item \inlinesection{Independence in audit tool use and protection from retaliation.} Power dynamics between auditors and the audited have a critical impact on accountability \cite{raji_outsider_2022, birhane_ai_2024}. Participants had audit results blocked from publication (P1) or unduly restricted in scope (P13) due to interventions from audit targets. Audit target retaliation and censorship was raised as a risk to both \emph{internal} auditors (P5), who face the threat of firings, social dismissal or professional demotion ~\cite{widder_its_2023, boag_tech_2022}, and \emph{external} auditors (P7), who face the threat of legal action under existing privacy and anti-hacking laws~\cite{urman_right_2024,raji_outsider_2022}. HCI research has explored software engineers' attempts to act on ethics concerns in the face of similar risks \citep{widder_its_2023, tahaei_privacy_2021}; further work could explore auditors' experiences specifically. Policymakers and stakeholders can take steps to provide protections for auditors through legal reforms~\citep{longpre_safe_2024} or legal funds such as the Coalition for Independent Technology Research.

\end{itemize}

\subsection{Moving beyond \emph{ad hoc} toolkits, towards shared infrastructure}
Participants agreed on the need for shared infrastructure that supports the auditing process. As one participant said, tools are \inlinequote{a superpower for journalists, and something that really is the future of accountability... There really needs to be some sort of public infrastructure [for auditing]}{7}. But developing high quality audit tools---even when adapting existing open source tools---took resources, and participants noted a lack of long-term investment: \inlinequote{We need more funding for this space... especially when it comes to infrastructure}{20}.

Currently, much of the funding for even open source audit tools comes from private, for-profit organizations. %
In our landscape analysis, we found that even the free tools currently dominating the audit tooling landscape were often built by large, for-profit tech companies (Fig.~\ref{fig:profit-count-second})---for example, {\profitNumTransp} of the {\nTransp} Transparency Infrastructure tools we found were built by for-profit organizations.
Audit tools like these can hold great power over the audit process: \blockquote{%
    Our determination of the performance of the algorithm carries a lot of weight within the organization. It's not like somebody else could just throw it and be like ``Oh, we'll go ask somebody else''... because we built the infrastructure, so we have that lever...
    If you can control the data sources or the ways to integrate algorithms into the data sources, that gives you power. %
}{3}
The external auditors we interviewed were especially skeptical of the data provided from these tools and aware of their unreliability, citing the shutdowns of transparency infrastructures like Reddit and Twitter's APIs (P24) as well as Facebook's CrowdTangle tool (P12). Open source tools (e.g., for data scraping) can sometimes fill the gap, but do not always cover the scope and complexity of practical audit work.

Our participants' aspirations for tooling envision another path---a path towards lasting public infrastructure that gives auditors additional levers to hold model operators accountable \citep{marda_public_2024}. Directions for HCI research and development include:

\begin{itemize}

\item \inlinesection{Tool catalogs \& other shared infrastructure.}
Tool selection was a major source of uncertainty for practitioners; as one expert suggested, \inlinequote{it's not just about creating a multitude of auditing tools but also about fostering decision support frameworks that empower practitioners to make informed choices based on the context they are dealing with}{25}. In addition to HCI work evaluating the efficacy of audit tools \citep{lee_landscape_2021, deng_exploring_2022, wong_seeing_2023, kaye_risky_2023, berman_scoping_2024}, future research could explore frameworks that assist audit practitioners in identifying and choosing between tools at each stage in an audit. Policymakers can invest in these frameworks and publish catalogs of vetted tools for auditors to reference, similar to the OECD's Tools for Trustworthy AI list \citep{oecd_tools_2021}. In general, shared AI inventories, AI incident databases, AI audit report registries, tool catalogs, regulatory guidance, and other centralized repositories or common transparency infrastructure could increase awareness, accessibility, and knowledge sharing, particularly if the audit community and affected stakeholders are empowered to not only utilize this infrastructure but also contribute to its development.

\item \inlinesection{Institutionalized tool maintenance \& funding.}
Currently, the burden of building and maintaining tools to address technical debt falls on the auditors, raising concerns about the sustainability of auditing efforts.
One participant suggested putting expiration dates on tools to avoid future inaccuracies (P7).
They also hoped to find \inlinequote{funders who are on board for longer tool maintenance projects}{7}, but another noted that attempts to raise support by connecting \inlinequote{performance issues and ML to downstream business KPIs}{4} was difficult when talking about \inlinequote{nebulous things, like fairness and bias}{4}.
Policymakers and foundations should set aside funding and resources for long-term tooling projects to ensure that audit tools are high quality, long-lasting, and address the wide range of needs identified in this study. Some initiatives, such as the Mozilla Technology Fund \citep{noauthor_mozilla_2024}, the U.K. AI Safety Institute's Systemic AI Safety Grants \citep{noauthor_fast_2024}, and the French AI Action Summit Public Interest fund \citep{ai_action_summit_public_2024} are already positioned to make these investments.
Researchers and practitioners should include long-term maintenance plans with newly developed tools, possibly in collaboration with civil society organizations such as the Linux Foundation, which hosts several of the open source tools we found (IBM's AI Fairness 360 \citep{bellamy_ai_2018}, for example).

\end{itemize}

\subsection{Ongoing Impact}
This work was completed as part of the Open Source Audit Tooling (OAT) project at the Mozilla Foundation and has already had demonstrable impact in policy engagement and funding. After submitting public comments, these findings were cited several times in the U.S. National Telecommunication and Information Administration (NTIA) ``Artificial Intelligence Accountability Policy Report'' \citep{goodman_ntia_2024}, the ``Summary report on the call for evidence on the Delegated Regulation on data access'' for the E.U. Digital Services Act \citep{leerssen_digital_2023}, and the U.K. AI Safety Institute's ``International AI Safety Report'' \citep{bengio_international_2025}. OAT team members presented these findings to regulators at the U.K.'s OfCom and the U.S. Federal Trade Commission. OAT team members also participated in advising the selection of two rounds of \href{https://foundation.mozilla.org/en/what-we-fund/programs/mozilla-technology-fund-mtf/}{Mozilla Technology Fund (MTF)} awardees, including 5 teams in an inaugural cohort (2022), and 8 teams in a follow-up cohort (2023) focused on AI audit tooling.

\section{Conclusion}

Ideally, AI audit studies will translate into tangible outcomes of accountability, but this outcome is far from certain. In order for the audit process to truly be feasible and effective, we---researchers, policymakers, and audit practitioners---need to invest in the infrastructure required for accountability. This will require a full effort on multiple fronts, including everything from the design and development of new tools; to new community infrastructure, communication standard-setting; to considering advocacy for certain policy positions. We cannot accept the minimum from AI auditing---we must push the boundaries of this practice until it becomes the meaningful mechanism of accountability it has the potential to be. %

\begin{acks}
    Thanks to our participants for volunteering their time and insight. Thanks also to seminar participants at Carnegie Mellon University and the Northeast HCI Meeting for their feedback on an earlier version of this work. This work was supported by the Mozilla Foundation. A.B. is supported by Science Foundation Ireland via the ADAPT Centre of Digital Content Technology funded under the European Regional Development Fund (ERDF) through Grant \#13/RC/2106\_P2. B.V. was supported by the University of Notre Dame IBM Tech-Ethics Lab.
\end{acks}

\ifbiblatex
\printbibliography
\else
\bibliographystyle{ACM-Reference-Format}
\bibliography{ryan-bibtex}
\fi

\appendix
\renewcommand\thefigure{\thesection.\arabic{figure}}
\setcounter{figure}{0}
\renewcommand\thetable{\thesection.\arabic{table}}
\setcounter{table}{0}

\section{Reflections}
\label{app:reflexivity}
We consider our own cultural and professional perspectives throughout the interviews and our analysis \citep{charmaz_constructing_2014}. In particular, we view our position as both external to prominent AI developers and deployments, but still situated within the Western AI industry as a critical consideration. Our project was financially supported by a prominent U.S.-based foundation, and all the authors are either graduate students or graduates of well-funded universities in the U.S. and Europe. 
We thus recognize that our analysis is primarily scoped to the U.S. and the E.U. and may not be representative of the global AI auditing landscape or appropriate for or informed by other contexts.
Because we drew our initial sources from our own fieldwork as audit practitioners and used English search engines for theoretical sampling, our dataset consists primarily of English language tools from from Western organizations, and our taxonomy reflects our own particular position.\footnote{We also attempted to run translated queries in non-English search engines (e.g., Baidu) and to add regional keywords (e.g., African) to our searches, but we were unable to find any additional tools with initial tests of these methods. It thus seems likely that non-English, non-Western tools exist that our theoretical sampling was unable to identify.}
Also, although we intentionally defined tools as resources more broadly and attempted to leave space for non-technical solutions in our analysis and discussion, we struggled to avoid a techno-solutionist framing in our conclusions \citep{wong_seeing_2023}. We hope that future work will expand and look outside of the mostly technical, Western perspective emphasized in this work.

\section{Glossary}
\label{app:glossary}

\newcommand{\defAccountability}{}
\begin{definition}
    \label{def:accountability}
    \textbf{Accountability:}
    \citet{bovens_analysing_2007} defines accountability as ``a relationship between an actor and a forum, in which the actor has an obligation to explain and to justify his or her conduct, the forum can pose questions and pass judgment, and the actor may face consequences.'' We use the term ``accountability'' in this legal-political sense, meaning to face consequential judgment for system behaviors and impacts that do not align with articulated expectations and standards.
\end{definition}

\begin{definition}
    \label{def:audit}
    \textbf{AI Audit:} \citet{birhane_ai_2024} define an audit as {\defAudit}. Independence, as outlined by \citet{birhane_ai_2024}, ensures the auditors (which may or may not belong to the same organization as the developers) are operationally distinct from the team that engineered the examined AI system, maintaining separation from the engineering process. Identified audit targets refer to concrete, specific objects of examination, ideally tied to real-world AI deployments or widely-used open-source algorithms or datasets, which serve as relevant proxies. Implicit or explicit accountability refers to audits designed to inform consequential judgments, measuring the deployment's behaviors and impacts against clearly articulated expectations.
\end{definition}

\begin{definition}
    \label{def:auditor}
    \textbf{AI Auditor:} An entity executing an AI audit, which may be viewed as either internal or external to the organization developing and/or operating the audited system. (Note that auditor independence is a nuanced spectrum, and there is not always a sharp divide between internal and external auditing.
    \begin{itemize}
        \item \textbf{Internal Auditor:} \citet{raji_outsider_2022} define an internal auditor as an entity executing an audit or investigation with some contractual relationship with the audit target. They typically seek to minimize corporate liability and test for compliance with corporate or industry-wide expectations. Internal auditors are often hired voluntarily or to meet regulatory mandates.
       \item  \textbf{External Auditor:} \citet{raji_outsider_2022} define an external auditor as an entity executing an audit or investigation without any contractual relationship with the audit target. They typically execute audits voluntarily with a broader mandate of identifying and minimizing the harm impacting their constituents.
    \end{itemize}
\end{definition}

\begin{definition}
    \label{def:audit-tool}
    \textbf{AI Audit Tool:} We use the term AI audit tool to refer broadly to {\defAuditTool}. Audit tools include resources that support algorithmic analysis and inspection (e.g., benchmarks/datasets, documentation templates) as well as resources that support the assessment of internal and external expectations for institutions across stages of design and development.

\end{definition}

\begin{definition}
    \label{def:abandonment-disgorgement}
    \textbf{Abandonment \& disgorgement:} In the context of algorithmic systems and regulatory enforcement, the terms algorithm abandonment \citep{johnson_fall_2024} and algorithm disgorgement \citep{li_algorithmic_2022} refer to the destruction or abandonment of a system deemed harmful, unethical, or in violation of societal or regulatory standards. Abandonment refers to ``an organization's decision to stop designing, developing, or using an algorithmic system due to its (potential) harms'' \citep{johnson_fall_2024}, while disgorgement further requires the deletion of both improperly obtained data and any machine learning models, algorithms, or outputs derived from such data \citep{li_algorithmic_2022}.
\end{definition}

\section{Additional Methods}
\subsection{Initial Sources}
\label{app:initial-sources}
We drew our initial list of {\nToolsInitial} tools from:
\begin{itemize}
    \item tools mentioned by our interviewees;
    \item tools mentioned in previous surveys of fairness and other toolkits \citep{deng_exploring_2022, lee_landscape_2021, hickock_ethical_2023}
    \item academic papers presenting new tools surfaced in a recent literature review of audit studies \citep{birhane_ai_2024}; this literature review included audit studies from \begin{itemize}
        \item the last five years of conference proceedings from: FAccT, AIES, EAAMO, AAAI, CVPR, ICWSM, WWW, WACV, EECV, and the ACL Anthology; also, the ACM Digital Library (including CHI and IC2S2) with the terms ``audit'', ``accountability'', ``case study'', ``bias'', ``fairness'', or ``assurance'' in the title, terms commonly used in AI studies published in computing venues;
        \item reports from the government agencies ICO and NIST
    \end{itemize}
    \item a convenience sample (accounting for {\nToolsSeed} of the initial {\nToolsInitial} tools) of other prominent audit tooling projects, academic papers, and tool-building organizations that we had encountered in our work as researchers and audit practitioners; most of these tools also appear in the sources above.
\end{itemize}

We included both tools designed specifically for AI auditing (such as AI fairness toolkits) and generic tools that have been used in AI audits (such as Selenium \citep{noauthor_selenium_2023}). Our dataset is not an exhaustive list of all tools that have been or could be used in AI audit practice. Note that there are substitutes and competitors for many of the tools in our dataset (such as qualitative coding software), but we did not include them unless we observed them in one of the sources above or our subsequent searches (\S~\ref{app:methods-tool-searching}). Conversely, inclusion of a tool does not indicate the authors' endorsement or preference.

\subsection{Theoretical Sampling}
\label{app:methods-tool-searching}
Category descriptors from the initial taxonomy that were sourced either from our labels or from descriptors used by tools already collected were then searched in combination with general keywords we identified that were commonly used in the algorithmic auditing domain. These keywords include the following: “tool”, “audit”, “algorithm”, “accountability”, “responsible AI”, “fairness”, “discrimination”, and “AI”. For each of our categories, we conducted English Google searches using a selection of one or more relevant keywords combined with each of our initial category descriptors (e.g., “participatory audit”, “participatory tool”, etc.). In order to determine a point of saturation for each search, we examined each Google search page of 30 results at a time until two pages in a row contained no references to specific audit tools. It is important to note that we conducted searches for all categories in our initial taxonomy, but focused first on less saturated categories (for example, harms discovery and tools using participatory methods) in order to provide well-rounded definitions of categories that were less common and/or visible. Categories with a more saturated set of examples were not given an equal amount of additional sourcing.

In addition to these targeted Google searches, we also added several new sources of tools to our initial list based on our initial taxonomy: \begin{itemize}
    \item News articles and reports by new organizations and civil society organizations including ProPublica, The Markup, the Pulitzer Center, the ACLU, and AlgorithmWatch;
    \item the Participatory ML Workshop at ICML \citep{kulynych_participatory_2020};
    \item an additional Google search for startups working on ``reg tech'' (regulatory tech).
\end{itemize}

\section{Interview Protocol}
\label{app:protocol}
\emph{We used the following protocol in each of our interviews. Because the interviews were semi-structured, not all participants answered every question in the same order. Bolded questions, however, were prioritized---we asked all these questions of nearly all participants. The rest were optional follow-ups.}

\noindent Thanks so much for taking the time to share your expertise. We really appreciate it and are looking forward to hearing your thoughts! [Briefly introduce the project.] [Confirm participant has completed consent form, remind participant of confidentiality, and confirm optional permissions.]

\textsc{Background}
\begin{itemize}
    \item \textbf{\textbf{How did you get involved in auditing to begin with? }}
    \item \textbf{\textbf{What do you hope to achieve?}}
\begin{itemize}
        \item Would you describe yourself as an internal or external auditor?
        \item What system was the target of your audit?
        \item What was the motivation behind your audit work?
        \item What are some notable successes?
        \item Notable failures?
        \item What were the most difficult aspects of the audit? How were those challenges overcome?
        \item What were the easiest aspects of the audit?
\end{itemize}
    \item Who do you consider to be stakeholders and why?
    \item Tell me about the people who have a role in designing and executing the audits you’re involved with.
\end{itemize}

\textsc{Tool Usage and Development}
\begin{itemize}
    \item \textbf{\textbf{Is there a specific tool or method (or set of tools and methods) that you employ? How do you choose these tools? What parts of the audits did they assist with?}}
\begin{itemize}
        \item \textbf{\textbf{What pain points did you encounter while using the tools?}}
        \item \textbf{Who made the decision to use this tool?\textbackslash{}Why do you use this tool and not others?}
\end{itemize}
    \item What prompted you to use/develop a tool? What are the system behaviors that you worry about?
    \item \textbf{\textbf{When do you know when to develop a tool vs. use an existing one?}}
    \item Can you help me understand why these tools are helpful, from an ethical perspective?
    \item What is the intent of the tool/method used? Do you find that the way you’ve used the tools/methods aligns with those intents?
    \item Some people are trying to build more open-source audit methodologies and tools that are freely available to the community. Is this an important goal for your audit practice? Why or why not?
\end{itemize}

\textsc{Exploring Gaps \& Challenges in Tools}
\begin{itemize}
    \item \textbf{What common obstacles do you encounter while designing, building, performing, and communicating about audits/tooling?}
\textbf{For tools, are there particular challenges (i.e., around adoption, maintenance, and distribution) that we should be aware of?}
    \item \textbf{\textbf{Do you find that there are needs that are unmet with existing auditing tools? What are they? / Is there any tool you wish you had but didn’t?}}
Have the tools/frameworks you’ve developed/used revealed any of the system behaviors that you are worried about? If yes, which ones (and to what extent)? If not, why do you think it did not uncover anything?
    \item In your experience, what are common properties that existing auditing tools/methods try to assess?
\begin{itemize}
        \item Do you think existing tools/methods are successful at measuring them?
        \item Are there things it would be good to measure that current tools don’t capture?
\end{itemize}
    \item To what degree do you find that existing auditing formats \& methodologies are useful and impactful? Are there formats/methodologies that you would like to see or see more of?
    \item \textbf{We’d like to get a sense of how resource-intensive your tool(s)/methods are.} Would you be willing to talk about how much it cost to perform audits or develop tools? How many people were involved? And how long does it take? How hard was it to do the audit and how much did it cost you?
\end{itemize}
\textsc{Wrap-up}
\begin{itemize}
    \item Is there anything else you’d like to talk about? Do you have any questions for me?
    \item {[Confirm optional permissions again.]}
\end{itemize}

\section{Additional Landscape Analysis}
\label{app:landscape}

To analyze the qualities of tools across our taxonomy, we manually labeled each tool with several tags describing the tool's documentation and function: license (open-source or proprietary); organization type (for-profit, non-profit, government, or academic); intended audit target (automated decision system, online platforms, large pre-trained online platforms autonomous vehicles, and/or other); intended user (internal and/or external); and format (e.g. API, software product, code/data repository, white paper, and/or other). One author created the labels and at least one other author reviewed each label for agreement.

We also supplemented our dataset with data from \href{http://www.crunchbase.com}{Crunchbase} \citep{noauthor_crunchbase_2023} accessed in September 2023, a platform for tracking funding, employment, revenue and other data for technology ventures, and \href{https://github.com/}{Github}, a platform for hosting and developing software. From Github, we scraped  repository activity---primarily the number of forks, stars, and issues---for the 98 tools with Github repositories in our dataset. Of the {\nOrgs} organizations in our dataset, we were able to access Crunchbase records for {\nOrgsCB}  (accounting for {\nToolsCB} tools); {\nOrgsCBEmployees} of these records include estimated employee counts. Of the {\nOrgsCBFirms} entries that are not for universities or government agencies, {\nOrgsCBFirmsRev} include revenue estimates. We also collected total venture funding (adjusted to U.S. dollars) for {\nOrgsCBFirmsPrivateFunding} firms out of the {\nOrgsCBFirmsPrivate} firms that are still private (i.e., have not undertaken an initial public offering). Additionally, we used the Google Scholar API to annotate each academic reference of an identified tool with the most recent available citation count in September 2023.

\begin{figure*}[p]
  \centering
  \includegraphics[width=0.6\linewidth]{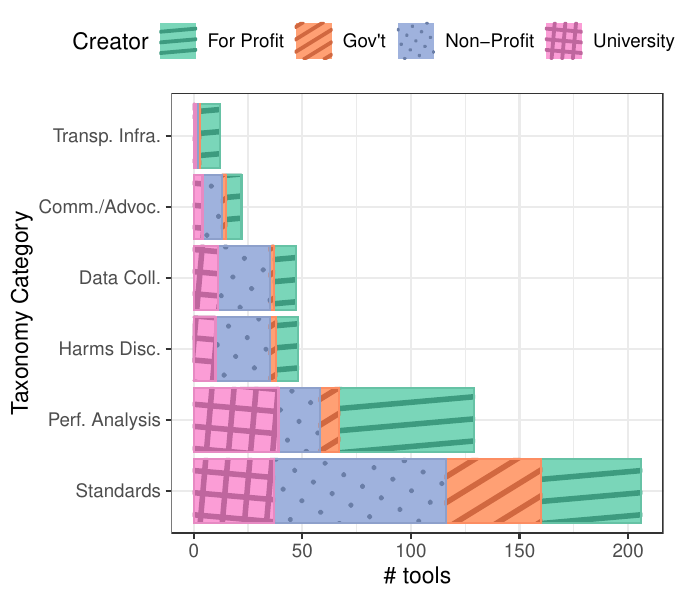}
  \caption{Number of tools by taxonomy category, sorted by type of organization (our classification). \taxonomyDisclaimer{}}
  \Description{In this bar graph, we present the number of tools categorized by taxonomy. Additionally, we've organized the data based on the type of organization associated with each tool. The graph visually represents the quantity of tools within each taxonomy category, with bars that are further divided or stratified based on their respective organizations.}
  \label{fig:profit-count}
\end{figure*}

\begin{figure*}[p]
  \centering
  \includegraphics[width=0.6\linewidth]{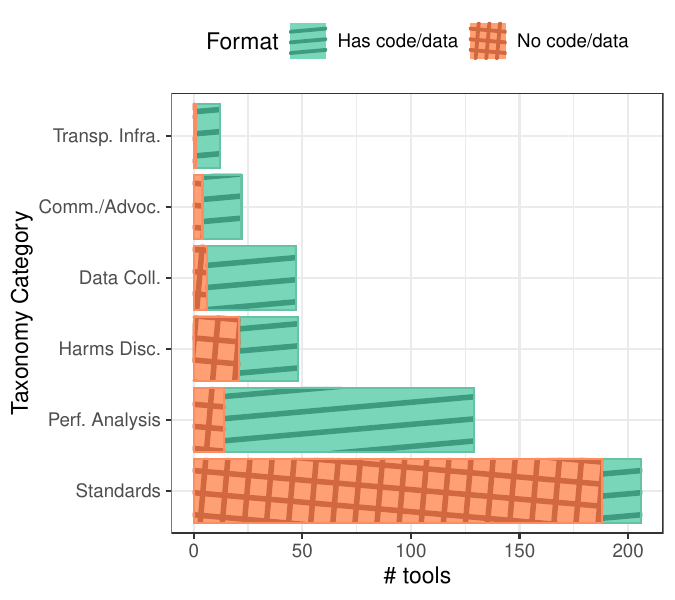}\\
  \caption{Number of tools with code in each taxonomy stage. \taxonomyDisclaimer{}}
  \Description{Bar graph showing the number of tools in each category based on the formats 'Has code/data' and 'No code/data'.}
  \label{fig:format-count}
\end{figure*}

\begin{figure*}[p]
  \centering
  \includegraphics[width=0.95\linewidth]{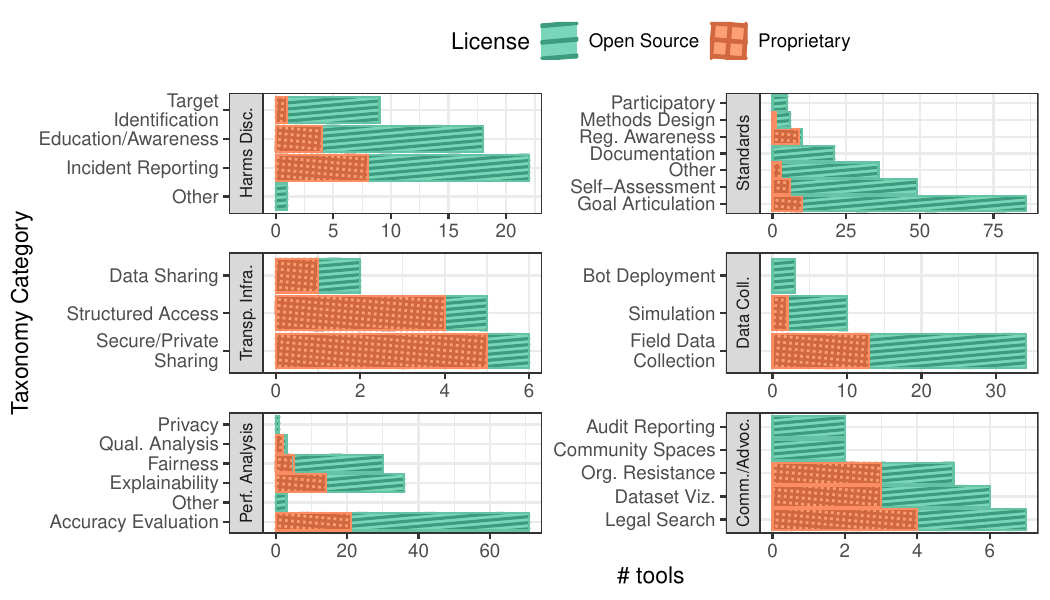}
  \caption{Number of tools by taxonomy category sorted by license type. \taxonomyDisclaimer{}}
  \Description{In this bar graph, we present the number of tools categorized by taxonomy. Additionally, we've organized the data based on the type of license associated with each tool. The graph visually represents the quantity of tools within each taxonomy category, with bars that are further divided or stratified based on their respective licenses.}
  \label{fig:license-count-second}
\end{figure*}

\begin{figure*}[p]
  \centering
  \includegraphics[width=0.95\linewidth]{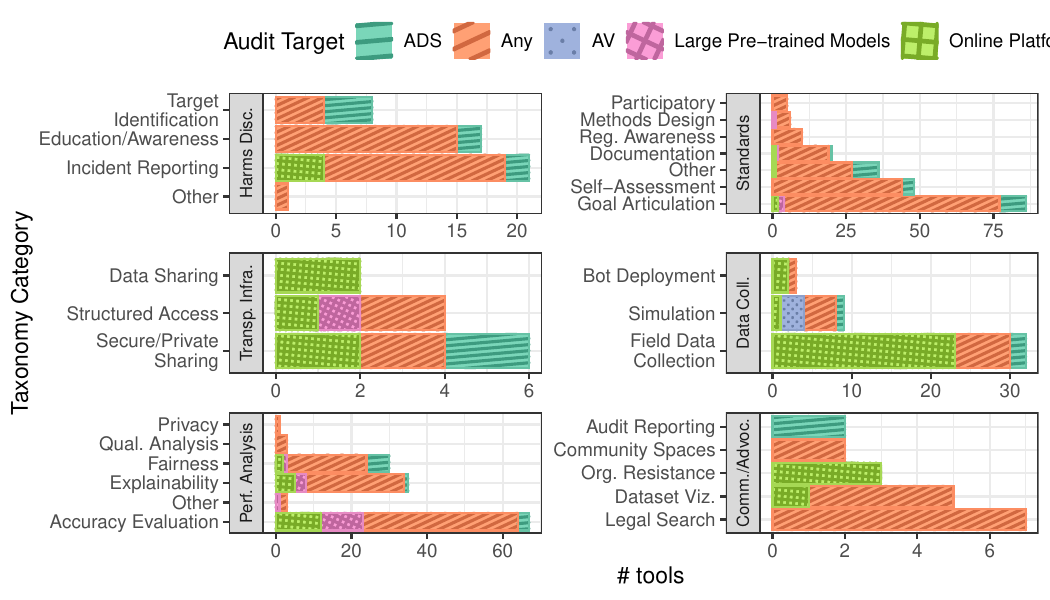}
  \caption{Number of tools by taxonomy category sorted by audit target. \taxonomyDisclaimer{}}
  \Description{In this bar graph, we present the number of tools categorized by taxonomy. Additionally, we've organized the data based on the type of audit target associated with each tool. The graph visually represents the quantity of tools within each taxonomy category, with bars that are further divided or stratified based on their respective audit targets.}
  \label{fig:target-count-second}
\end{figure*}

\begin{figure*}[p]
  \centering
  \includegraphics[width=0.95\linewidth]{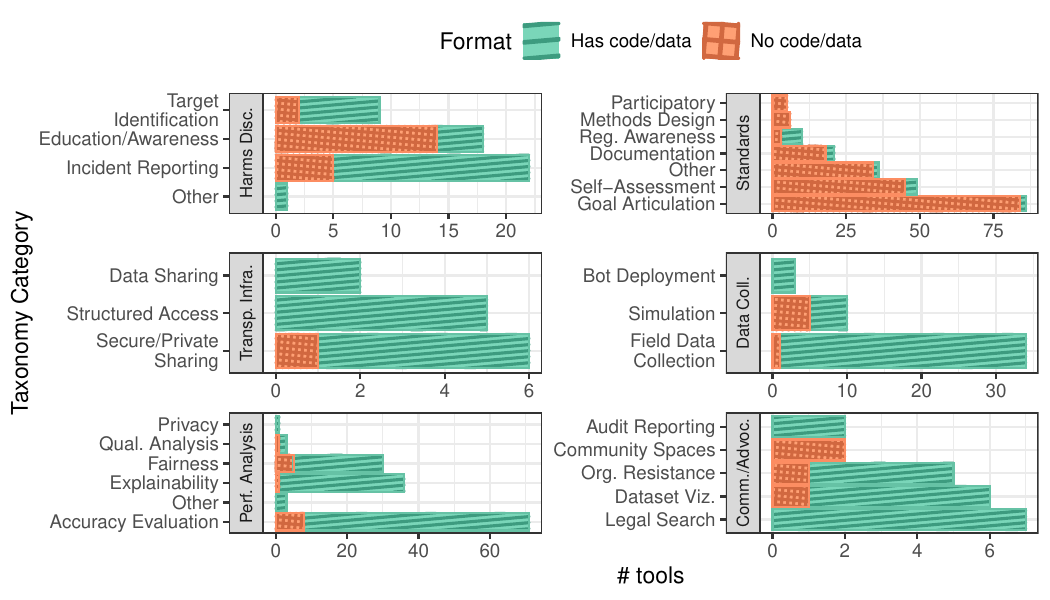}\\
  \includegraphics[width=0.95\linewidth]{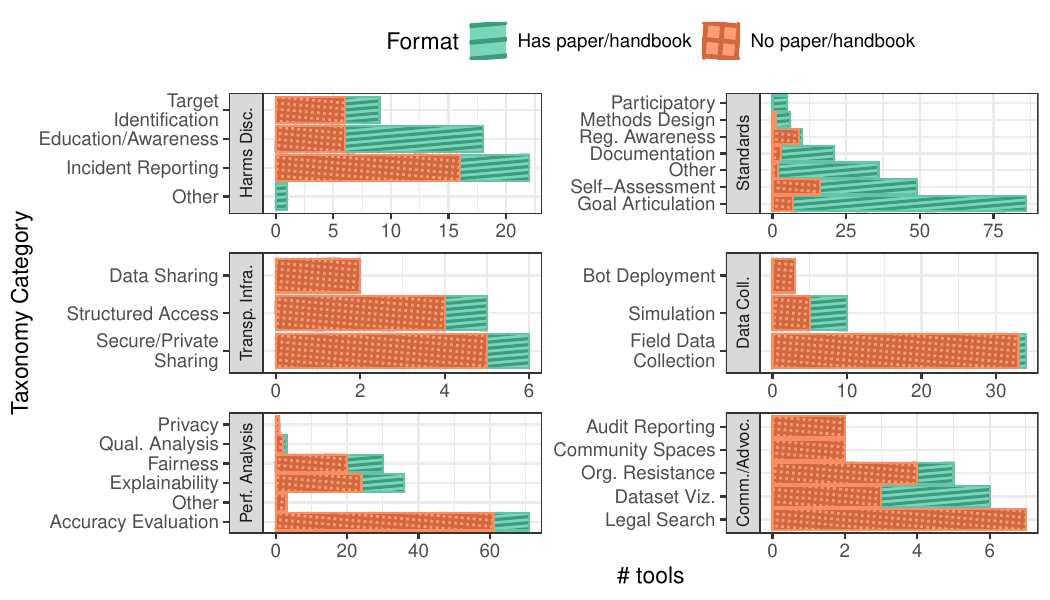}
  \caption{Number of tools by taxonomy category sorted by format. \taxonomyDisclaimer{}}
  \Description{In this bar graph, we present the number of tools categorized by taxonomy. Additionally, we've organized the data based on the type of format associated with each tool. The graph visually represents the quantity of tools within each taxonomy category, with bars that are further divided or stratified based on their respective formats.}
  \label{fig:format-count-second}
\end{figure*}

\begin{figure*}[p]
 \centering
 \includegraphics[width=0.8\linewidth]{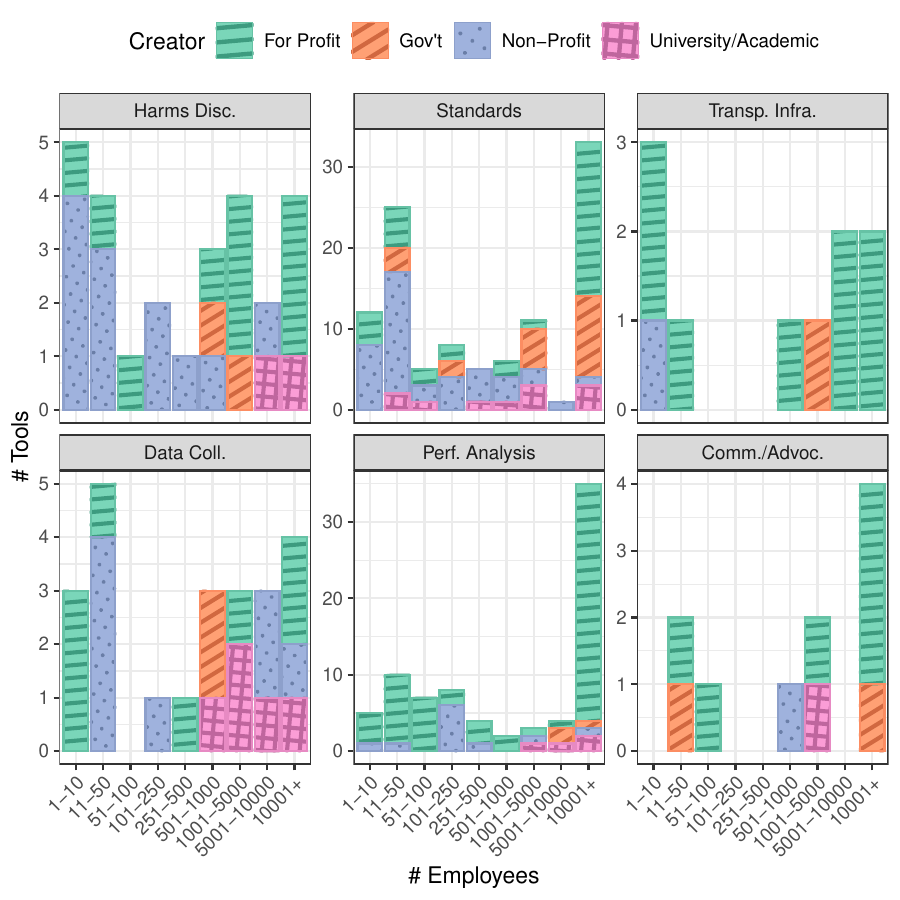}
 \caption{Number of tools by taxonomy category. \taxonomyDisclaimer{} Workforce size of creating organization sourced from \href{http://www.crunchbase.com}{Crunchbase} \citep{noauthor_crunchbase_2023}. Sorted by type of organization (our classification). Tools from organizations without Crunchbase entries excluded.}
 \Description{In the provided bar chart, we display the number of tools categorized under six major taxonomy categories. These categories are further divided based on their organization type, which can belong to one of four groups: 'For profit,' 'Government,' 'Non Profit,' or 'University/Academic.' The chart visually represents the quantity of tools in each of these categories using vertical bars, allowing readers to understand the distribution of tools across different taxonomies and organizational types.}
 \label{fig:profit-employees}
\end{figure*}

\begin{figure*}[p]
  \centering
  \includegraphics[width=0.8\linewidth]{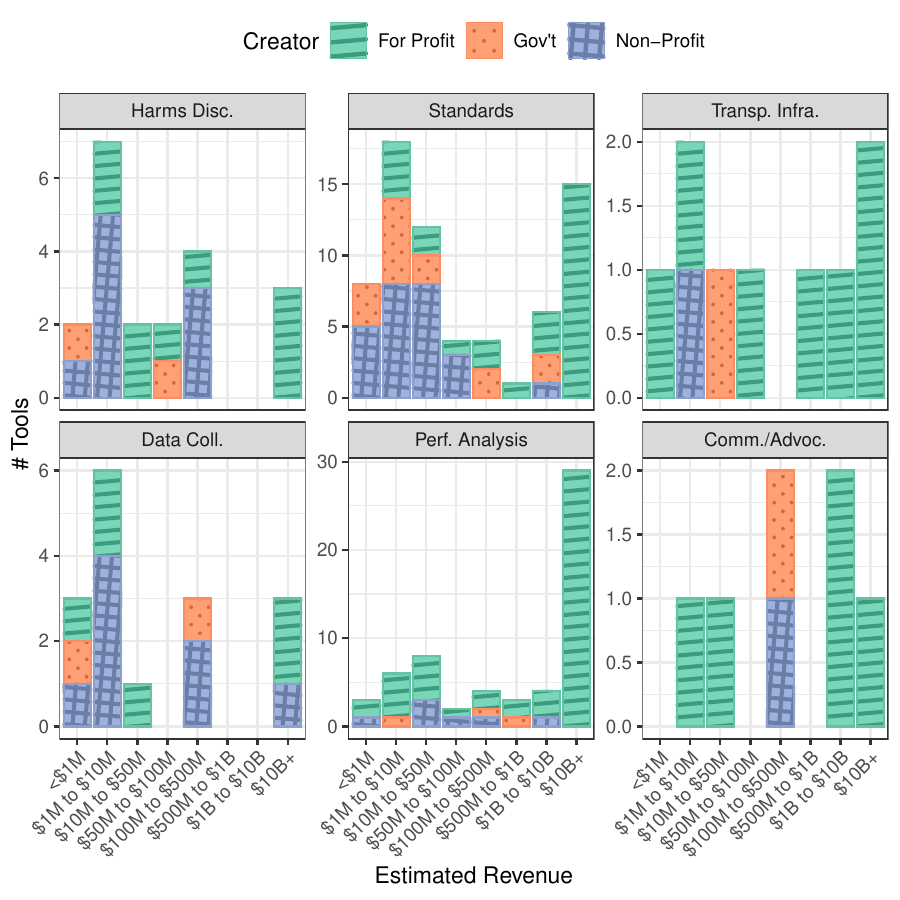}
  \caption{Number of tools by taxonomy category. \taxonomyDisclaimer{} Estimated revenue of creating organization sourced from \href{http://www.crunchbase.com}{Crunchbase} \citep{noauthor_crunchbase_2023}. Tools from organizations without Crunchbase revenue estimates excluded.}
  \Description{In this bar graph, we present the number of tools categorized by taxonomy. Additionally, we've organized the data based on the type of organization and estimated revenue associated with each tool. The graph visually represents the quantity of tools within each taxonomy category, with bars that are further divided or stratified based on their respective type of organization and estimated revenue.}
  \label{fig:profit-revenue}
\end{figure*}

\begin{figure*}[p]
  \centering
  \includegraphics[width=\linewidth]{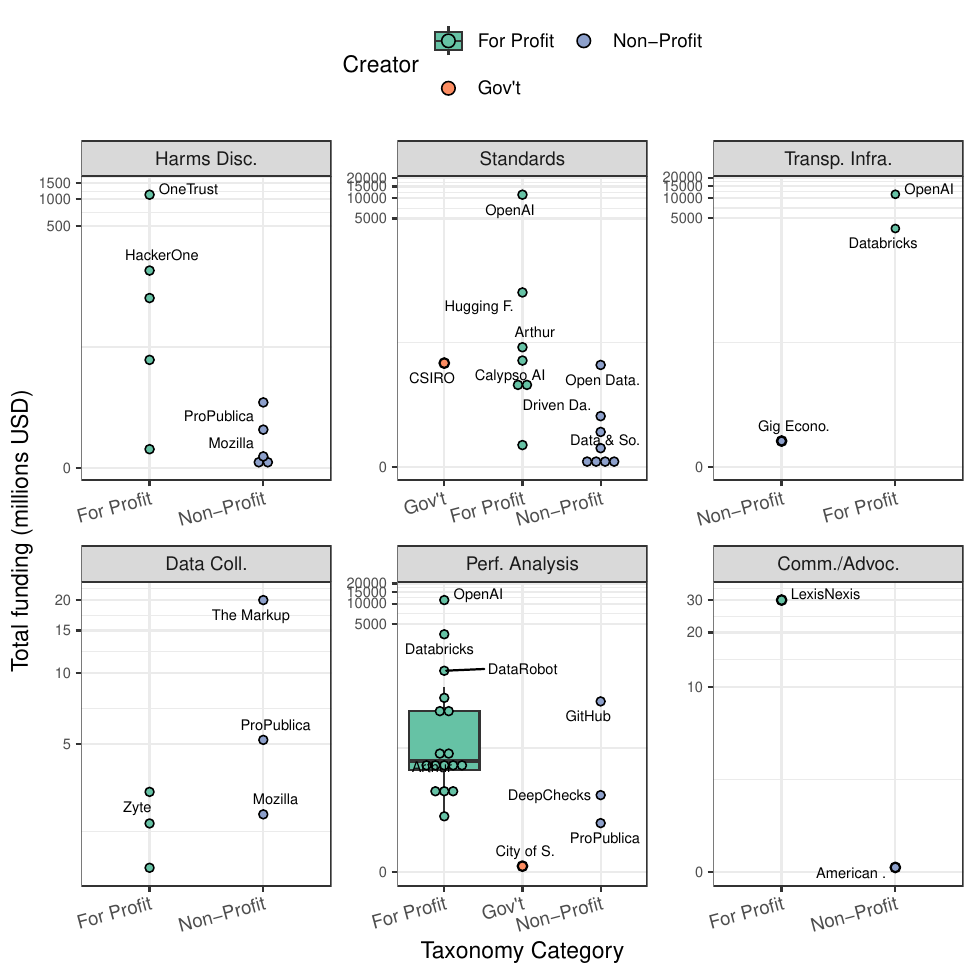}
 \caption{Private (pre-IPO) organizations with \href{http://www.crunchbase.com}{Crunchbase} entries ($114/311$ organizations). Total funding sourced from Crunchbase \citep{noauthor_crunchbase_2023}. \taxonomyDisclaimer{}}
  \Description{ Boxplot showing distribution of tools by organization type and total funding in millions USD.}
  \label{fig:profit-funding}
\end{figure*}

\begin{table*}[!htbp] \centering 
  \caption{Top-funded audit tool builders, per \href{http://www.crunchbase.com/}{Crunchbase} data \citep{noauthor_crunchbase_2023}. Includes only private (pre-IPO) organizations with Crunchbase entries.} 
  \label{tab:top-funding} 
  \centering
\begin{tabular}{@{\extracolsep{5pt}} cccccc} 
\\[-1.8ex]\hline 
\hline \\[-1.8ex] 
Organization & Total funding (millions USD) & Estimated revenue & Employees & Stages \\ 
\hline \\[-1.8ex] 
\href{https://www.crunchbase.com/organization/openai}{OpenAI} & 11303.12 & \$50M to \$100M & 501-1000 & Perf. Analysis, Transp. Infra., Standards \\ 
\href{https://www.crunchbase.com/organization/databricks}{Databricks} & 3497 & \$500M to \$1B & 5001-10000 & Transp. Infra., Perf. Analysis \\ 
\href{https://www.crunchbase.com/organization/onetrust}{OneTrust} & 1120 & \$100M to \$500M & 1001-5000 & Harms Disc. \\ 
\href{https://www.crunchbase.com/organization/datarobot}{DataRobot} & 1000.598 & \$100M to \$500M & 501-1000 & Perf. Analysis \\ 
\href{https://www.crunchbase.com/organization/hugging-face}{Hugging Face} & 395.2 &  & 101-250 & Perf. Analysis, Standards \\ 
\href{https://www.crunchbase.com/organization/github}{GitHub} & 350 & \$100M to \$500M & 1001-5000 & Perf. Analysis \\ 
\href{https://www.crunchbase.com/organization/h2o-2}{H2O.ai} & 251.099999 & \$10M to \$50M & 251-500 & Perf. Analysis \\ 
\href{https://www.crunchbase.com/organization/weights-biases}{Weights \& Biases} & 250 & \$10M to \$50M & 251-500 & Perf. Analysis \\ 
\href{https://www.crunchbase.com/organization/hackerone}{HackerOne} & 159.4 & \$10M to \$50M & 1001-5000 & Harms Disc. \\ 
\href{https://www.crunchbase.com/organization/bugcrowd}{Bugcrowd} & 78.65 & \$1M to \$10M & 501-1000 & Harms Disc. \\ 
\href{https://www.crunchbase.com/organization/arthur-ai}{Arthur} & 60.3 &  & 51-100 & Standards, Perf. Analysis \\ 
\href{https://www.crunchbase.com/organization/pymetrics}{Pymetrics} & 56.63 & \$1M to \$10M & 51-100 & Perf. Analysis \\ 
\href{https://www.crunchbase.com/organization/fiddler-labs}{Fiddler AI} & 45.2 & \$1M to \$10M & 11-50 & Perf. Analysis \\ 
\href{https://www.crunchbase.com/organization/truera}{TruEra} & 42.284998 &  & 51-100 & Perf. Analysis \\ 
\href{https://www.crunchbase.com/organization/cognitivescale}{CognitiveScale} & 40 & \$10M to \$50M & 51-100 & Perf. Analysis \\ 
\href{https://www.crunchbase.com/organization/calypso-ai}{Calypso AI} & 38.2 & \$500M to \$1B & 11-50 & Standards, Perf. Analysis \\ 
\href{https://www.crunchbase.com/organization/seldon}{Seldon} & 33.691771 & \$1M to \$10M & 51-100 & Perf. Analysis \\ 
\href{https://www.crunchbase.com/organization/open-data-institute}{Open Data Institute} & 32.835579 & \$1M to \$10M & 11-50 & Standards \\ 
\href{https://www.crunchbase.com/organization/lexisnexis}{LexisNexis} & 30 & \$1B to \$10B & 10001+ & Advocacy \\ 
\href{https://www.crunchbase.com/organization/the-markup}{The Markup} & 20 & \$1M to \$10M & 11-50 & Data Coll. \\ 
\hline \\[-1.8ex] 
\end{tabular} 
\end{table*}

\begin{table*}[!htbp] \centering 
  \caption{20 most popular Github repositories for tools in our database, sorted by number of forks.} 
  \label{tab:top-repos} 
\begin{tabular}{@{\extracolsep{5pt}} ccccc} 
\\[-1.8ex]\hline 
\hline \\[-1.8ex] 
Tool & Forks & Issues & Stars & Stages \\ 
\hline \\[-1.8ex] 
\href{https://github.com/scrapy/scrapy}{Scrapy} & 10458 & 667 & 52280 & Data Coll. \\ 
\href{https://github.com/SeleniumHQ/selenium}{Selenium} & 8101 & 242 & 30129 & Data Coll. \\ 
\href{https://github.com/appium/appium}{Appium} & 6052 & 137 & 18618 & Data Coll. \\ 
\href{https://github.com/carla-simulator/carla}{CARLA} & 3562 & 1079 & 11075 & Data Coll. \\ 
\href{https://github.com/slundberg/shap}{SHAP} & 3245 & 806 & 22426 & Perf. Analysis \\ 
\href{https://github.com/openai/evals}{Evals} & 2568 & 130 & 14631 & Perf. Analysis \\ 
\href{https://github.com/marcotcr/lime}{LIME} & 1795 & 120 & 11492 & Perf. Analysis \\ 
\href{https://github.com/EleutherAI/lm-evaluation-harness}{Language Model Evaluation Harness} & 1673 & 338 & 6312 & Perf. Analysis \\ 
\href{https://github.com/Trusted-AI/adversarial-robustness-toolbox}{Adversarial Robustness Toolbox} & 1146 & 150 & 4738 & Perf. Analysis \\ 
\href{https://github.com/Trusted-AI/AIF360}{AI Fairness 360} & 827 & 199 & 2401 & Perf. Analysis \\ 
\href{https://github.com/seldonio/seldon-core}{Seldon Core} & 827 & 203 & 4334 & Perf. Analysis \\ 
\href{https://github.com/interpretml/interpret}{Interpret} & 726 & 105 & 6198 & Perf. Analysis \\ 
\href{https://github.com/google/BIG-bench}{Big Bench} & 582 & 107 & 2802 & Perf. Analysis \\ 
\href{https://github.com/tensorflow/privacy}{Tensorflow Privacy} & 447 & 121 & 1915 & Perf. Analysis \\ 
\href{https://github.com/bethgelab/foolbox}{Foolbox} & 421 & 27 & 2713 & Perf. Analysis \\ 
\href{https://github.com/fairlearn/fairlearn}{Fairlearn} & 410 & 163 & 1888 & Perf. Analysis \\ 
\href{https://github.com/facebookresearch/PurpleLlama}{Purple Llama} & 404 & 1 & 2476 & Perf. Analysis \\ 
\href{https://github.com/github/CodeSearchNet}{CodeSearchNet} & 384 & 14 & 2172 & Perf. Analysis \\ 
\href{https://github.com/pair-code/lit}{Language Interpretability Tool} & 351 & 113 & 3453 & Perf. Analysis \\ 
\href{https://github.com/microsoft/responsible-ai-toolbox}{Error Analysis} & 344 & 86 & 1325 & Perf. Analysis \\ 
\href{https://github.com/microsoft/responsible-ai-toolbox}{Responsible AI Toolbox} & 344 & 86 & 1325 & Perf. Analysis \\ 
\hline \\[-1.8ex] 
\end{tabular} 
\end{table*} 

\begin{figure*}[p]
   \centering
   \includegraphics[width=\textwidth]{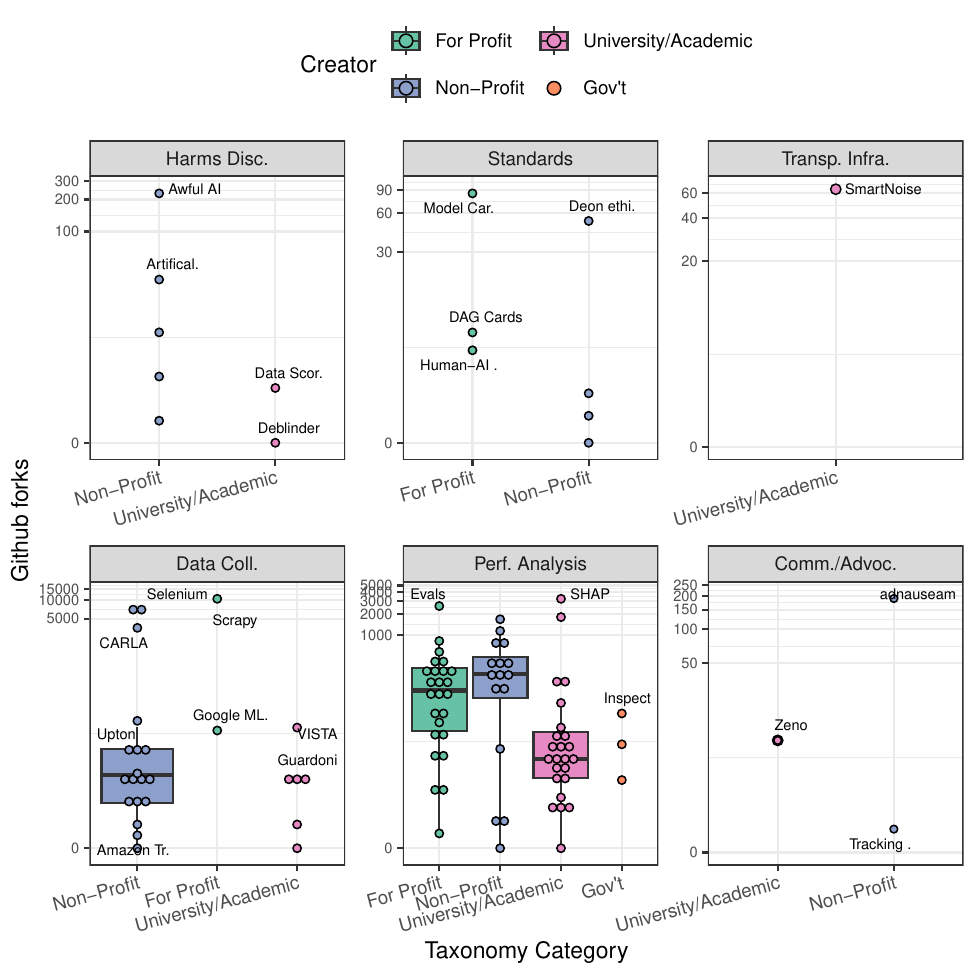}
   \caption{Github forks by taxonomy category (for tools with Github repositories), sorted by type of organization (our classification). \taxonomyDisclaimer{} Box-and-whisker plots included for categories with more than 10 points.}    
   \Description{In a boxplot graph, we present the distribution of Github forks by taxonomy category. We stratify each boxplot by the organization type of its creator and label outliers.}
   \label{fig:profit-usage-forks}
\end{figure*}

\begin{figure*}[p]
   \centering
   \includegraphics[width=\textwidth]{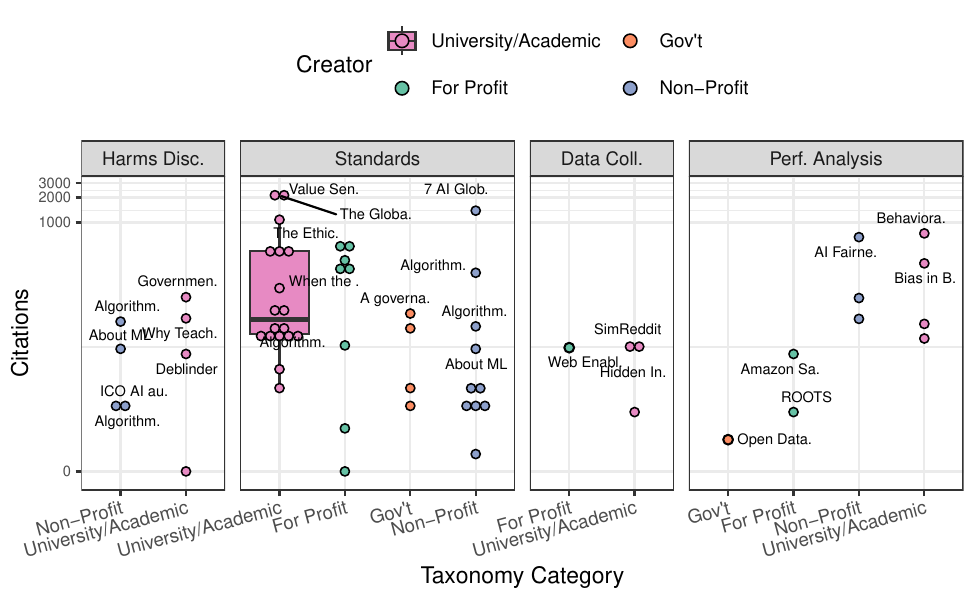}
   \caption{Citations by taxonomy category (for tools with papers), sorted by type of organization (our classification). \taxonomyDisclaimer{} Box-and-whisker plots included for categories with more than 10 points.}    
   \Description{In a boxplot graph, we present the distribution of citations by taxonomy category. We stratify each boxplot by the organization type of its creator and label outliers.}
   \label{fig:profit-usage-citations}
\end{figure*}

\end{document}